\documentclass[journal]{IEEEtran}
\IEEEoverridecommandlockouts

\usepackage{cite}
\usepackage{url}
\usepackage{balance}
\usepackage{adjustbox}
\usepackage{subfigure}
\usepackage{tabularx} % For full width table
\usepackage{pifont}% http://ctan.org/pkg/pifont
\usepackage{amsmath}
\usepackage{array}
\usepackage{multirow}
\usepackage{rotating}
\usepackage{xcolor}
\PassOptionsToPackage{table,xcdraw}{xcolor}



\usepackage{caption}

\newcolumntype{L}[1]{>{\raggedright\let\newline\\\arraybackslash\hspace{0pt}}m{#1}}
\newcolumntype{C}[1]{>{\centering\let\newline\\\arraybackslash\hspace{0pt}}m{#1}}
\newcolumntype{R}[1]{>{\raggedleft\let\newline\\\arraybackslash\hspace{0pt}}m{#1}}

\newcommand{\ie}{{\em i.e.,} }
\newcommand{\eg}{{\em e.g.,} }
\newcommand{\etc}{{\em etc.,} }

\newenvironment{sitemize}{%
  \begin{list}{$\bullet$}{%
    \setlength{\itemsep}{0.1cm}%
    \setlength{\leftmargin}{1.0em}%
    \setlength{\topsep}{0.1cm}%
    \setlength{\parsep}{0mm}}%
  }{\end{list}}

\usepackage{bbding}
\usepackage{dblfloatfix}
\usepackage{color, colortbl}
\usepackage{graphicx}
\usepackage{amssymb}% http://ctan.org/pkg/amssymb
\usepackage{pifont}% http://ctan.org/pkg/pifont
\usepackage{algorithm}
\usepackage[noend]{algpseudocode}
\usepackage[nolist]{acronym}
\usepackage{fancyvrb}
\usepackage{listings}

\usepackage{amsthm}
\newtheorem{thm}{Theorem}%[section]

\newtheorem{lem}[thm]{Lemma}

\newtheorem{defn}[thm]{Definition}

\def\equationautorefname~#1\null{%
  (#1)\null
}

\ifCLASSINFOpdf
\else
\fi

\usepackage[indentfirst=false]{quoting}
\quotingsetup{vskip=0pt}
\newcounter{myquote_counter}
\newsavebox{\myquotebox}
\newsavebox{\myquoterefbox}
\definecolor{quotebackcolor}{rgb}{0.9,0.9,1}
\newlength{\quotewidth}
\newlength{\quoterefindent}
\newcommand*{\scaledquotefactor}{0.95}
\newenvironment{myquote}[1]%
{\renewcommand*{\scaledquotefactor}{1.0}%
   \begin{lrbox}{\myquoterefbox}\begin{minipage}[b]{0.8\quotewidth}%
       \em #1%
   \end{minipage}%
   \end{lrbox}%
   \begin{lrbox}{\myquotebox}\begin{minipage}[b]{1.0\quotewidth} 
}%
{\end{minipage}\end{lrbox}%
\begin{center}%
 \mbox{\begin{minipage}{\quotewidth}
                        \scalebox{\scaledquotefactor}{\usebox{\myquotebox}}\\[1pt]
          \hspace*{\quoterefindent}\scalebox{\scaledquotefactor}{\usebox{\myquoterefbox}}
       \end{minipage}
      }      
\end{center}%
\par}

\graphicspath{{.}
{Figures/}
}

\begin{document}
%enable this switch line numbering on
%\linenumbers
\title{ForestFirewalls: Getting Firewall Configuration Right in Critical Networks (Technical Report)}
%{Identifying the Missing Aspects of the ANSI/ISA Best Practices for Security Policy \\ (Extended Case Studies)}

%TODO: update this before submitting!
%\author{Dinesha~Ranathunga,~\IEEEmembership{Student Member,~IEEE,}
%        Matthew~Roughan,~\IEEEmembership{Senior Member,~IEEE,}
%        Paul~Tune,~\IEEEmembership{Member,~IEEE,}       
%        Phil~Kernick and 
%        Nick~Falkner }% <-this % stops a space

\author{\IEEEauthorblockN{Dinesha Ranathunga\IEEEauthorrefmark{1},
Matthew Roughan\IEEEauthorrefmark{1},
Paul~Tune\IEEEauthorrefmark{1},
Phil Kernick\IEEEauthorrefmark{2} and
Nick Falkner\IEEEauthorrefmark{3}\\}
\IEEEauthorblockA{\IEEEauthorrefmark{1}ACEMS, University of Adelaide, Australia\\}
\IEEEauthorblockA{\IEEEauthorrefmark{2}CQR Consulting, Australia\\}
\IEEEauthorblockA{\IEEEauthorrefmark{3}School of Computer Science, University of Adelaide, Australia\\
Email: \{dinesha.ranathunga, matthew.roughan, paul.tune, nick.falkner\}@adelaide.edu.au\\
          phil.kernick@cqr.com}}

%\author{Paper \#, X pages}

\maketitle

\thispagestyle{plain}
\pagestyle{plain}

\begin{abstract}
Firewall configuration is critical, yet often conducted manually with inevitable errors, leaving networks vulnerable to cyber attack \cite{wool2004}. The impact of misconfigured firewalls can be catastrophic in \ac{SCADA} networks. These networks control the distributed assets of industrial systems such as power generation and water distribution systems. Automation can make designing firewall configurations less tedious and their deployment more reliable. 

In this paper, we propose \textit{ForestFirewalls}, a high-level approach to configuring \ac{SCADA} firewalls. Our goals are three-fold. We aim to: first, decouple implementation details from security policy design by abstracting the former; second, simplify policy design; and third, provide automated checks, pre and post-deployment, to guarantee configuration accuracy. We achieve these goals by automating the implementation of a policy to a network and by auto-validating each stage of the configuration process. We test our approach on a real SCADA network to demonstrate its effectiveness.
\end{abstract}


% keywords
\begin{IEEEkeywords}
firewall auto-configuration, SCADA network security, security policy, policy verification, Zone-Conduit model.
\end{IEEEkeywords}

\IEEEpeerreviewmaketitle

%%%%%%%%%%%%%%%%%%%%%%% Section 1 %%%%%%%%%%%%%%%%%%%%%%%%%%%%%%%%%%
\section{Introduction}

\begin{myquote}{Rubin and Greer~\cite{rubin1998}}
``The single most important factor of your firewall's security is
how you configure it.''
\end{myquote}

% SCADA networks are important for critical infrastructure, firewalls play a major role compared to IT networks
Supervisory Control and Data Acquisition (SCADA) networks control the distributed assets of many industrial
systems. Power generation and water distribution are just two examples
that illustrate the critical nature of these networks. Others include
factory automation, sewage management, airport control systems and
chemical plant control.

SCADA networks are not like corporate IT networks
\cite{stouffer2008}. IT networks can accept a degree of
reliability orders of magnitude lower than the network controlling a
power station. A fault in the latter will cost serious money,
if not lives.

At the same time, SCADA networks often incorporate highly vulnerable
devices. The \ac{PLCs} that control
physical devices such as gas valves have highly
constrained memory and computational power. Today, they often include
network functionality such as a TCP/IP stack, but exclude
sophisticated security functionality. 

Despite their name, \ac{PLCs} are not {\em user} programmable. The plant
operator does not program them -- that requires a programming board
that pushes low-level code into an EPROM. The devices come
pre-installed with code (and security holes). There are many PLCs in a
power station, along with similar devices providing telemetry, and their 
upgrade is likely to occur only during a major overhaul of a
plant, which might happen once in a decade (if that often).

These devices would be vulnerable if
exposed to the Internet, and a plant operator cannot fix the
vulnerabilities. Air gaps have been proposed as a
solution to protect these devices, but an air gap is no longer a
feasible approach for many reasons. In fact, Byres calls the idea a
myth~\cite{byres13:_air_gap} to emphasise how poor a solution it is.
The only viable protection today is a firewall, or series 
of firewalls \cite{stouffer2008, byres2005}.

\indent As Rubin and Greer note~\cite{rubin1998}, it is, therefore, vital that
these firewalls are configured correctly.  Misconfiguration of
\ac{SCADA} firewalls can lead to security breaches, resulting in
significant physical and environmental damage, financial loss or worse, the loss of human lives.
A recent example is the hacking of a German steel mill by cyber attackers in 2014 that destroyed its blast furnace \cite{bbc2014}.

Unfortunately, firewall configuration, in practice, is a complicated
and repetitive manual task.  It involves training in proprietary and
device-specific configuration languages, and long and complex device
configurations. Lack of automation tools to assist this task has resulted in
unoptimised, error-prone configurations \cite{wool2004, wool2010,
  ranathunga2015}.
% that often deviate from the industry recommended network security guidelines

The problem is exacerbated in SCADA plants where industrial engineers
generally lack specialised networking and security knowledge. Such
knowledge is often brought in through third party contractors. These
IT security specialists are usually unfamiliar with the
specific requirements of industrial engineering, and are on-site
only for brief periods.
%simple specification not enough - need checking: need assurance that generated configs fully reflect specified policy: only whats permitted in policy is allowed  through ebverything else is blocked

A cost-effective alternative to training plant engineers to become 
IT specialists is to build network operations tools that derive
firewall configurations from high-level policy. 
%SDN etc are potentinal long term solutions _WHY?? explain but we need something in the short term that works well
Approaches using SDN have been proposed
\cite{soule2014,levin2013}, but they remain a distant reality for
SCADA networks, where TCP is still a recent innovation. And power plants are
insecure now~\cite{ranathunga2015}! SCADA networks need a solution that
works now, using off-the-shelf technology. In this paper, we propose
such a solution: \emph{ForestFirewalls}.

Our system provides a mechanism for specification of security policy
at a level a non-IT specialist could understand. What's more, it
forces good designs on its users through principles derived
from the study of real SCADA firewall
configurations~\cite{ranathunga2015} and \ac{ISA} best practices
\cite{isa2007,byres2012,stouffer2008}. Most notably:
\begin{sitemize}

\item {\em Single source of truth:} more specifically,
  ``Security managers need a single place to look for the corporate
  policies on who gets in and who doesn't.'' \cite{howe1996}. This
  general principle in computer science \cite{bellovin2009} applies doubly here.
%This goal is much realised using our system in a SCADA context. 
% http://en.wikipedia.org/wiki/Single_Source_of_Truth

\item {\em Simplify:} we don't try to provide every possible security
  feature. At best, advanced features create confusion, and at
  worst, bad implementations can create security flaws.

\item {\em Verify everything again and again:} there is a clear danger
  in assuming any one piece of software functions correctly, from the
  firewall up to and including our own system. We check the
  configuration works at every level possible.

% \item {\em Inclusive policy zones:} it is tempting to circumvent
%   regionalization in the network to allow ``more specific'' policies,
%   or holes to specific servers. This creates an illusion of security,
%   not the fact. We do not allow such circumvention.

\item {\em No implicit rules:} implicit rules allow unexpected
  interactions, and undesirable consequences \cite{ranathunga2015}. Desired flows must be
  explicitly allowed.
  
\item {\em Rule order should not matter:} it should be possible to add, or subtract a
policy rule without considering its effect on every other rule. Surprisingly, none of the existing
firewall configuration platforms \cite{cisco2010,cisco2014,checkpoint2008,juniper2011} achieved this. 
Operators using these tools, must provide correct rules {\em and} maintain correct rule order to avoid adverse interactions.
%have to maintain correct rule order, ancillary to providing correct rules, to avoid unexpected interactions.

\item {\em Separate structure from function \cite{pearce1998}:} decoupling\\  
network topology (\ie structure) from policy specification (\ie function) facilitates high-level requirements based policies 
without network-centric minutiae like IP addresses.

\item {\em Convenience:} security and convenience are usually at odds,
  but wherever possible convenience should be provided. This is not a luxury
  -- lack of convenience is one of the main reasons operators
  circumvent {\em their own} security.
\end{sitemize}

Our system comprises a suite of tools to write policy, validate,
test configurations, and create real configurations, and is designed to be 
user extensible.

We demonstrate it with a real example, derived from the actual (but
anonymised) firewall configurations of a real SCADA plant. The
example is intentionally small for clarity, but it shows a core
set of functionality. The example includes several zones, two
firewalls, and multiple real services much as they would run in the
real network. Our testbed uses two different firewalls: one Cisco and
one Linux-based, in order to show both the device independent nature
of our policy language, and that heterogeneous network devices can be
configured in the same network.  The network offers multiple services:
DNS, HTTP, ... \etc and we use test traffic on the network to show correct
function. 

The proof of the pudding is that we can specify this network's policy 
in only 80 \ac{LoC}, to generate the equivalent of 2720
device-level LoC found in a real SCADA case study
\cite{ranathunga2015}. This order of magnitude reduction, along with
rigorous validation, shows the value of {\em ForestFirewalls}.

%%%%%%%%%%%%%%%%%%%%%%% Section 2 %%%%%%%%%%%%%%%%%%%%%%%%%%%%%%%%%%
\section{Related work}
% parallel work
% There are lots of firewall products and security and firewall management tools available on the market.
A useful firewall configuration platform should allow policies to be specified abstractly, flexibly enough and in detail.
There are many products and security management tools with varying levels of sophistication introduced by firewall vendors \cite{cisco2010,cisco2014,checkpoint2008,juniper2011},
but policies still cannot be specified using high-level requirements.

For example, Cisco introduced {\em security levels} for quick and easy access between internal and external firewall interfaces \cite{cisco2010}, 
but these cannot specify detailed traffic restrictions. Hence, \ac{ACLs} are required to supplement these levels. Security levels may also not map to clear security policies. This hinders firewall auto-configuration, which needs clear policies \cite{bellovin2009} to permit traffic. 

High-level security polices need to be based on security abstractions.
The choice of the abstraction determines the level of decoupling between network topology and security policy.
For one, Firmato \cite{bartal1999s} uses a role-based abstraction in its network grouping language that is independent of the 
firewalls and routers used in the network. But, the abstraction {\em does not intuitively map} policy to topology and requires minutiae 
like IP addresses to be input through the policy specification, to implement policy on a network instance.
%But, the abstraction requires minutiae such as IP addresses as input, when implementing policy on a network instance.

For another, Cisco has introduced security policy management tools (\eg VNMC for VSG policy management) to cater for complexities introduced in network virtualisation \cite{cisco2014b}. For scalability, the tools allow operating systems (\ie VMs) to be allocated to zones and policies to be defined per zone. 
But, each VM still needs to be defined using minutiae like hostnames. 
{\em ForestFirewalls} decouples policy from topology using a more intuitive {\em Zone-Conduit abstraction}.
The direct {\em zone-to-network} mapping allows policy to {\em exclude} IP addresses or hostnames as input.
%IP addresses are not required to be provided through the policy specification since 
%overcomes this problem by using a more intuitive {\em Zone-Conduit abstraction}.
%Similarly, Tesseract \cite{tesseract2007} allows direct control of Ethernet and IP based services,  but lacks the ability to abstract low-level policy configuration details.

% FANG, LUMETA
The ability to debug configuration errors is also a highly desirable feature in a firewall configuration platform.
But, this debugging should support abstract queries, free of network centric minutiae.
Fang \cite{mayer2000} and Lumeta \cite{wool2001} are interactive tools that facilitate firewall rule debugging, 
but the queries they support use network-centric minutiae as input.
%they rely on network-level queries that use minutiae detail as input.

%NETKAT
Also desirable would be if the configuration platform supported more general-purpose abstract queries like service reachability and traffic isolation queries. 
Network programming languages supporting such queries have been proposed (\eg NetKAT \cite{anderson2014}), 
but they too, rely on minutiae as input; and are not specifically aimed at configuring firewalls.

% PGA
Ability to create and correctly compose distributed policies is also necessary in a firewall configuration platform. Doing so, allows users from multiple policy sub-domains (\eg SCADA engineers, Corporate admins) to manage their own policies. Policy Graph Abstraction (PGA) \cite{prakash2015} provides such a framework for SDN networks, Cloud infrastructure and Network Functions Virtualisation (NFV) environments. It can automatically compose distributed policies into a coherent, conflict-free policy set. But, PGA leaves out how logs, alerts and alarms (\ie reports) should be enabled in security-critical middle boxes like firewalls.
These reports are critical to detect misconfigurations and network security setup failures. 
Our solution allows reports to be encapsulated with security policy via a framework designed by us.
%However, ?? - no formal verification?, 

Additionally, it is essential in SCADA networks to have assurance of expected configuration behaviour prior to deployment, since downtime must be minimised.
SANE \cite{sane2006} proposes a central Domain Controller with trusted privileges to reduce end-host initiated attacks in corporate networks. It supports
topology-independent high-level declarative policies but provides no pre-deployment guarantee of correct configuration behaviour. 
{\em ForestFirewalls} uses simple automated emulations to achieve this goal.

% TODO: Openstack - congress and group based policies 
%Network virtualisation requires traditional network boundaries to be broken, to allow N:1 mapping between operating systems (\ie VMs) and a network port. Cisco has introduced security policy management products (\eg VNMC for VSG policy management) to cater for the complexity this introduces to network management \cite{cisco2014b}. For scalability, the products allow VMs to be allocated to zones and  policies to be defined per zone. However, each VM still needs to be defined using low-level detail such as hostnames. 

% Tesseract
%Tesseract implements a network control plane that enables direct control of Ethernet and IP based services \cite{tesseract2007}. It promotes 
%centralised policy implementation, but lacks the ability to abstract low-level policy configuration details.

% SANE 
%SANE uses a central Domain Controller with trusted privileges \cite{sane2006} to reduce end-host initiated attacks in corporate networks. It supports
%topology-independent high-level declarative policies, but provides no assurance of expected configuration behaviour prior to deployment. Such assurance, 
%even via simple automated emulations, is essential in SCADA networks where downtime must be minimised.

Most related works do not propose a high-level description that intuitively decouples policy from topology. In some cases  \cite{bartal1999s}, topology needs to be explicitly mapped to policy through the specification per host/subnet basis. There is also no automated pre- and post-deployment verification of policies. Moreover, none of the above works examine \ac{SCADA} networks, with unique security requirements and best practices compared to Corporate networks. 

%There are some additional relevant works but we will discuss these in context.
The prior work also does not address a key practical issue- {\em complexity}. Firewall vendors have concentrated on new and impressive features to create systems with as much or more complexity as the base firewall configurations.

Our research tackles this problem head-on. Our solution, \textit{ForestFirewalls}, uses security abstractions to drastically reduce firewall policy specification complexity. These abstractions decouple policy from topology intuitively, so firewall policies can be described using
high-level requirements independent of vendor and network intricacies.
Automated verification is also built-in to our system to mitigate misconfigurations. 
{\em ForestFirewalls} also provides a formalism to compose rule sets into a coherent, conflict-free policy set.
These features collectively,  make firewall configuration a commodity skill rather than a specialisation, 
so business managers and plant engineers alike can manage their SCADA firewalls.

% Frenetic and Netkat
%The network programming language introduced in the Frenetic project \cite{foster2011}, is able to capture dynamic policies, but cannot assist with queries related to network reachability.
%NetKAT is a language \cite{anderson2014} built on a complete equational theory to address this shortfall. It additionally supports features such as traffic isolation and compiler correctness. But the language is not specifically aimed at configuring firewalls and does not provide means to generate filtering rules.

% PGA
%Policy Graph Abstraction (PGA) \cite{prakash2015} provides a framework to write distributed policies and automatically compose them into a coherent, conflict-free policy set.
%The solution enables reliable and automatic distributed policy management for SDN networks, Cloud infrastructure and Network Functions Virtualisation (NFV) environments.
%However, PGA leaves out how logs, alerts and alarms (\ie reports) should be enabled in security-critical middle boxes such as firewalls.
%Firewall reports are critical to detect security misconfigurations and failures in network security setup. 
%{\em ForestFirewalls} allows reports to be encapsulated with security policy through a reporting framework we have built.
% also decoupling of policy from network in PGA is non-intutive??
% no formal verification 

%%%%%%%%%%%%%%%%%%%%%%% Section 3 %%%%%%%%%%%%%%%%%%%%%%%%%%%%%%%%%%
\section{Requirements}
\label{sec:requirements}

Bush and Bellovin \cite {bellovin2009} identified the following core requirements of an automated security configuration system:
\begin{sitemize}
\item {\em Clear policies:} an automated system cannot resolve between a plausible and a correct policy \cite{bellovin2009}.  For example, between allowing HTTP access to a publicly shared Web server or to a sensitive internal Web server. So the policy must be clearly understood by a Manager.
\item {\em Database driven:} all device configurations and their changes must be recorded in a \textit{single reference point} \cite{bellovin2009}, which allows fast response to security incidents as well as accurate security audits.
\item {\em Meta-configurations:} specifications or instructions\\ about real configurations need to be obtained by abstraction and parameterisation.
\end{sitemize}
However, there are extra issues not described by \cite{bellovin2009}. For one, there  is an assumption that the auto-configuration system generating the configurations is correct. But proof of correctness must stem from validation, as we describe next.

\subsection {Policy verification}

\begin{figure}[t!]
%\center
\captionsetup{aboveskip=8pt}
\centerline{\includegraphics[scale=0.32]{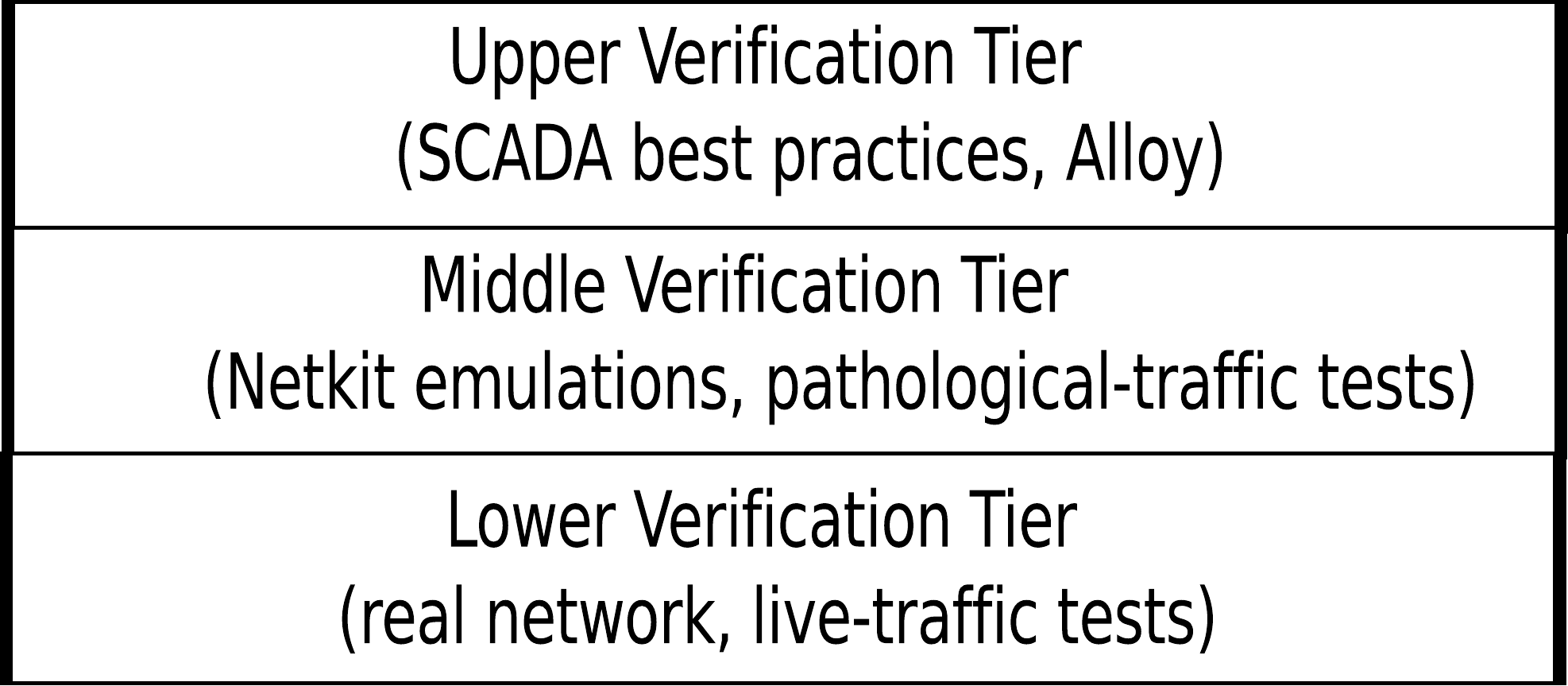}}
\caption{Policy verification tiers.}
\label{fig:layered}
%\vspace{-3mm}
\end{figure}

SCADA operators need assurance that the device configurations generated produce the expected security outcome, both pre- and post-deployment. We use multiple verification stages 
(\ref{fig:layered}) to provide this assurance.

\smallskip
\noindent \textit{\textbf{Upper verification tier}}: 
it is important initially to check a specified firewall policy against available industry best practices \cite{byres2005, stouffer2008}. These automated compliancy tests are critical in SCADA networks where more restrictive practices are required to minimise their vulnerability to threats from less secure networks (\eg the Internet).
Best practice violations can be accurately identified by conducting {\em equivalence} and {\em inclusion} checks on the canonicalised policies. See~\ref{sec:verification}. 
%Direct violations of best practices indicate exploitable vulnerabilities of the network implementing the policy, and should be prevented. 

Complex firewall policies also produce unintended consequences through rule overlaps \cite{yuan2006,liu2008}.  So it is additionally necessary to check policies from high-level through firewall-level for inconsistencies. We do so accurately, using a mathematical and logic based formal tool: \textit{Alloy} \cite{jackson2012}.

\smallskip
\noindent
\textit{\textbf{Middle verification tier}}: the second stage helps debug configuration problems prior to deployment. Network emulation offers a cost effective method to test configurations before actual deployment \cite{knight2013}. The \textit{Netkit} open source software package \cite{pizzonia2008} provides such an emulation platform with virtual devices and interconnections via \ac{UML}. Automated pathological traffic tests, together with Netkit emulations, can verify that the generated configurations produce the expected outcome prior to deployment.

\smallskip
\noindent
\textit{\textbf{Lower verification tier}}: the final stage guarantees that the real firewalls operate as intended, post deployment. The automated tests are extended from emulations to the real network, to generate live-traffic and reveal unexpected configuration behaviour in the real firewalls.

\smallskip
\noindent Automated verification can drastically reduce the number of firewall misconfigurations. It can be used to identify best-practice violations and adverse policy interactions, and prevent those from propagating to the physical firewalls. Most importantly, it provides users with a guarantee of the security outcome prior to implementation. 
%\vspace{-4mm}
%This is particularly useful in successfully responding to intrusions and attacks.

\subsection{Firewall reporting}
A second issue not described by \cite{bellovin2009}, but key to autoconfiguration, is the need to be able to constantly examine the status of security provided by the firewalls in a network. We discuss this briefly here to keep within the scope of this paper -- see \cite{ranathunga2015T} for an extended discussion.

% why do we need to consider reporting?
Firewall security status can be monitored using the diverse logs, alerts and alarms available. We refer to these together as {\em firewall reports}. These reports are critical in identifying firewall misconfigurations and detecting failures in the network security setup. But, the industry standards available for reporting and analysing firewall data are scant and vague. So, we developed a firewall reporting framework \cite{ranathunga2015T} considering the {\em granularity} and {\em scope} requirements of the various reporting use cases.

%We made several key findings, in developing this reporting framework.
We learnt key lessons, in developing  this reporting framework.
For one, a firewall is a poor source of data in some reporting use cases (\eg accounting) \cite{ranathunga2015T}, although they are often used in this way. 
In those cases, it is better to report elsewhere in the network and release a firewall's resources for its primary function: {\em traffic filtering}. 
Where firewall reporting {\em must} be enabled, it needs to be at the correct granularity. 
Employing an {\em inferior granularity} prevents collecting sufficient data to satisfy use-case requirements, using an {\em overly-fine granularity} simply wastes resources. 

Additionally, security policy and reporting policy need to be {\em encapsulated} (\ie coupled) together \cite{ranathunga2015T}. 
Decoupling the two can lead to bad decisions, for instance, it would allow adding security policies that are not verified.

\subsection{Decouple policy from network}
A third issue not described by \cite{bellovin2009}, but key to autoconfiguration, is the need to decouple policy from the network implementation.

% policy should NOT need to know about underlying network details.. it should be indep and strictly depend on policy maker decision ONLY
% current approach to firewall configuration is too complexed and non-scalable..need to decouple policy from network
In practice, network  architects and business managers decide what type of services are allowed through firewalls. Network engineers then implement these policies. 
Intuitively,  separation of the network intricacies from policy specification better suits these distinct phases. Conceptually this is
analogous to the separation of architects and building contractors in construction. Contractors don't usually decide what roof shape a building should have.

Network topology can change often in response to new business needs, upgrades and service demands. 
This may alter the devices and administratively assigned parameters such as IP addresses and hostnames in the network.

Comparatively, security policies are static. These policies commonly only involve 
dozens of distinct services \cite{ranathunga2015}, so policy complexity is relatively low to that of the network. 
This relative simplicity and invariant nature leads to decoupling policy from the network. 
The adage \textit{``Structure and function should be independent''} \cite{pearce1998} truly applies here. 

\noindent Decoupling \textit{structure} from \textit{function}, has these advantages:
\begin{sitemize}
\item {\em Policy is specifiable via high-level, vendor neutral\footnote{Vendor platform and device independent requirements.} requirements:} which assists management-level policy makers not fluent in network-centric details.
\item {\em It centralises policy management and promotes reuse:} \\ a single topology-independent organisation policy (\ie \textit{source of truth}) can be maintained across sites.
\item {\em It streamlines network changes and upgrades:} the policy can be quickly re-mapped to a new topology, retaining previous levels of protection.
%\item {\em Simplifies security audits:} the complexity of policy checks is reduced in the absence of network-level details such as IP addresses.
\item{\em It simplifies best-practice enforcement:} best practice standards can be {\em precisely} specified in absence of proprietary details of specific networks.
\end{sitemize}

\noindent Security abstractions are key to decoupling policy from the underlying network.

%%%%%%%%%%%%%%%%%%%%%%% Section 4 %%%%%%%%%%%%%%%%%%%%%%%%%%%%%%%%%%
\section{Security abstractions}

% what are the features of a good high-level security abstraction
A good high-level security abstraction captures the underlying network security concepts naturally and concisely. For one, in real networks, we might group
systems with a similar set of traffic services enabled. For another, traffic restrictions between two systems may be enforced by a single or a series of firewalls. A good abstraction should decouple \textit{what} is restricted between systems from \textit{how} it is restricted.
 
% ISA provides a good alternative 
The \ac{ANSI}/ \ac{ISA} introduce the \textit{Zone-Conduit abstraction} to segment and isolate the sub-systems in a 
control system \cite{isa2007}.  But, the Zone-Conduit model in its original specification is too flexible for automation. To increase its precision, we needed to add several extensions \cite{ranathunga2015}.  We describe our modified version here.
\vspace{-2mm}

\subsection{Our modifications to the \ac{ISA} model}

The Zone-Conduit model is a graph $G=(Z,C)$ where $Z$ is the set of zones, and $C \subset Z \times Z$ is the set of conduits. 

A \textit{zone} is a logical or physical grouping of an organisation's systems with similar security requirements, based on criticality and 
consequence \cite{isa2007}. By grouping systems in this manner, a \textit{single zone-policy} can be defined for all members of a zone. For example, 
3 disjoint security zones can be defined to accommodate low, medium and high-risk systems, with each device assigned to its respective zone, 
based on their security level needed. A low-risk system can be accommodated within a medium or high security zone without compromising 
security, but not vice versa. 

A \textit{conduit} provides the secure communication path between two zones, enforcing the policy between them \cite{isa2007}. 
Security mitigation mechanisms (\eg firewalls) are implemented within a conduit. A conduit could consist of multiple links and firewalls, but logically is a single connector.
Conduits abstract \textit{how} a policy is enforced, so we can focus on \textit{what} needs to be enforced.

% the ISA model is not precise enough so we added extensions
We conducted real SCADA firewall configuration case studies \cite{ranathunga2015} and found that the ISA Zone-Conduit model in its original specification is {\em too flexible} for 
automation. For one, the ISA model allows alternate ways of defining zones and conduits to cater for business models. It loosely permits 1:n or n:1 mapping between conduits, firewalls and policy. 
 
We refine the model through several extensions \cite{ranathunga2015}.  First, we enforce a 1:1 mapping between policies and conduits. Second, dedicated
Firewall-Zones are used to capture firewall management policies. Third, Abstract-Zones are used to capture the distinct policy requirements of serial firewalls. Carrier-Zones are also used to abstract any carrier based transit outside of an administrative domain's control. 

The ISA standard also loosely allows sub-zones to be defined, enabling multiple policies within a zone.
Doing so, implies that selected subsystems in a zone (\eg a server) could have their own separate policies. 
Allowing exceptions would impart a false sense of security: these systems are only as secure as the zone itself, in the absence of firewalls enforcing a real separation. 
We tighten the specification further in our approach by strictly enforcing a \textit{single zone-policy}. 
  
A \textit{single zone-policy} leads to every device within a zone having the same set of permissions to initiate, accept or block one or more services. 
Hence, a zone is the smallest unit of abstraction for which a policy can be applied to and we do so simply and unambiguously using \textit{inter-zone flows}. \\

\noindent Once extended, the best practice produces a tight specification suitable for auto-configuration.
We can now begin to formally define firewall security policies based on this refined Zone-Conduit model.

\subsection{Policy on a single conduit}

A {\em conduit policy} in the refined Zone-Conduit model can be constructed from an ordered set of rules $[p_1, p_2, ..., p_n]$ that act on packet sequences to accept, deny, or in some cases, modify them.
% formal definition of a firewall policy rule
Particularly, a policy rule $p_1$ operates on  ${\cal A} = \{ atomic \ packet \ sequences\}$, where an atomic packet sequence is a {\em complete packet sequence} that cannot be decomposed into smaller subsequences (except themselves and the null sequence- $\phi$). 
 {\em Complete} means that a decision on two concatenated complete subsequences is the same as that on the joint sequence, \ie $ p_1( s_1 + s_2 ) = p_1(s_1) + p_1(s_2) \text{,} \forall s_1, s_2 \in {\cal A}$, where $+$ is an associative {\em concatenation operator}. ${\cal A}$ is closed under $+$ and $\phi$ is the identity.  
 So, ${\cal A}$ is a {\em monoid} and policy rules $p(\cdot)$ are monoid endomorphisms on ${\cal A}$ (\ie mappings from ${\cal A}$ to itself that preserve the semigroup structure of the operator $+$ and identity $\phi$).
% (they preserve the identity because $p(s + \phi) =p(s) + p(\phi) = p(s)$ and hence $p(\phi) = \phi$).

% TODO: do we talk about timing being a crucial component of the packet sequences?
Typical policy rules accept/deny packets, \ie for $A \in {\cal A}$
\begin{equation}
  \label{eq:policy}
  p_{A}(s)= 
  \begin{cases}
    s,    & \text{if } s \in A,  \ // \ accept \\
    \phi, & \text{if } s \in A^c,  // \ deny.
  \end{cases}
\end{equation} 

% don't consider rules that modify packets
This type of rule doesn't allow modification or creation of packets. 
Real firewall rules can modify or create
packets. For instance a firewall might
\begin{sitemize}
\item update certain header fields related to QoS; or
\item might defragment, or fragment packets; or  
\item be integrated with Network Address Translation (NAT) or 
Virtual Private Network (VPN) functionality.
\end{sitemize}

The scope for packet modification is huge, but within a firewall, many
modifications don't change fields that would affect further rules in subsequent firewalls, 
\eg QoS or TTL changes. In order to
have a tractable problem, we restrict the firewall rules to such
modifications, and thus consider all rules to be in the form given in
\ref{eq:policy}. 

% rules cannot be defined for ANY atomic packet sequence- firewall technology limits it
We also cannot construct a policy rule for any possible subset of ${\cal A}$, due to the limitations of technology used in a firewall.
The subsets of ${\cal A}$ for which rules can be defined is actually a  {\em sigma algebra}
$\sigma({\cal A})$ \cite{patrick1995}.

The particular sigma algebra will be generated by the finest possible partition of ${\cal A}$ determined by the firewall technology used.
So for a given firewall technology, ${\cal A}$ can be broken into sets $A_i \subset {\cal A}$ such that $A_i \cap A_j = \{\}$
and $\cup_i A_i = {\cal A}$, and we can implement rules $p_{A_i}$, but cannot define any rule $p_B$ where $B$ is a strict subset of some $A_i$. 

% rules are actually implemented using predicates and actions
Firewall rules in practice however, are implemented by specifying a predicate and an action. A firewall enforces actions based on predicate matching, 
and firewall policies found in practice are built using multiple policy rules \cite{ranathunga2015}.

\subsection{Positive, explicit policies}

A firewall policy in practice, is made up of multiple predicate matching rules that can be combined using several strategies: {\em first match}, {\em last match} or {\em all match}.
If we presume here the conservative security option: an implicit \verb|deny-| \verb|all| rule in place, then an accept rule $q^{a}_{m}$ (where $m$ is the predicate) defines accept packet set $A = \{s
\in {\cal A} \mid s \prec m \}$, where $s \prec m$ denotes $s$ matches $m$. A single
deny rule $q^{d}_{m}$ has no affect, \ie $A = \{\}$.

When we combine two (ordered) rules $(q^{t}_{m_1}, q^{t}_{m_2})$ where $t \in \{a,d\}$,
we define operators based on the matching order:
\begin{enumerate}
\item first match
  \[ q^{t}_{m_1} \oslash q^{t}_{m_2} =       
      \begin{cases}
        q^{t}_{m_1}, & \text{if } s \prec m_1 \\
        q^{t}_{m_2}, & \text{if } s \prec m_2 \; \text{ and} \; s \not\prec m_1 \\
        deny,         & \text{otherwise,}\
      \end{cases}   \label{eq:oslash} \\  
   \]

\item last match 
  \[ q^{t}_{m_1} \ominus q^{t}_{m_2} =       
      \begin{cases}
        q^{t}_{m_2}, & \text{if } s \prec m_2 \\
        q^{t}_{m_1}, & \text{if } s \prec m_1 \; \text{ and} \; s \not\prec m_2 \\
        deny,         & \text{otherwise.}\
      \end{cases}   \label{eq:oslash} \\  
   \]
\end{enumerate}

Clearly the operations $\oslash$ and $\ominus$ are
associative, \eg $(q_1 \oslash q_2)\oslash q_3=q_1 \oslash (q_2
\oslash q_3)$, but not commutative or equivalent. 

This is a problem. We need a policy to hold the same semantics regardless of how the rules are ordered, in order to simplify policy specification.

So, we restrict ourselves to {\tt accept rules}, conditional on an
implicit {\tt deny-all} rule. This restriction
\begin{sitemize}
\item is rich enough to represent all rules \cite{ranathunga2015P} and
\item the operators are then commutative and equivalent \cite{ranathunga2015P}.
\end{sitemize}

Equivalent results also hold if we only allow deny rules, but this option
is less secure as it is easier to accidentally leave something out
of a deny list than to include it.

Therefore, we adopt a security whitelisting model, \ie we restrict \textit{inter-zone flows} to express positive abilities\footnote{Refers to the ability to initiate or accept a traffic service.} and {\em deny all} flows that are not explicitly allowed. Doing so, renders the rule order irrelevant in a policy. 
A policy now holds the same semantics, irrespective of how its rules are organised. Hence, policy makers need not consider
the order when adding or removing policy rules. By being explicit, we also prevent services being accidentally enabled implicitly. 

We can use the conduit-policy definition to also check for {\em equivalence} as well as {\em inclusion} of policies \cite{ranathunga2015P}.
In \ref{sec:verification} we will demonstrate how doing enables evaluation of actual firewall policies against industry best practices for violations.

\smallskip
\noindent So far, we considered policy rules on a single conduit. We now generalise these policies to a network, or rather the simplified Zone-Conduit model of the network.

\subsection{Network policy}

For policy space $\Phi = \{ p : {\cal A} \rightarrow {\cal A} \mid p \mbox{ is a monoid endomorphism} \}$, 
we can define a network policy as
\begin{defn}[Network Policy]
  A network policy $\mathcal{P} = (G,P)$ means a Zone-Conduit graph
  $G(Z,C)$ with policy functions $p_{ij} \in \Phi$ for $(i,j) \in C$.
  \label{def3}
\end{defn}
% can also define semantics partitions
When comparing network policies, it is also useful to partition the policies on the networks 
(\eg $p_{ij}$ in Definition \ref{def3}) into equivalence classes \cite{ranathunga2015P}.
We refer to these partitions as {\em Semantic Partitions (SPs)} as they have the same meaning. 
%Doing so, we extend the notion of comparison of two network policies to the comparison of their SPs.

\subsection{Isolation of traffic}
% why do we need traffic isolation guarantees?
An important element in architecting a network's security is the proper isolation of traffic between zones,
\ie it is necessary to guarantee the privacy and integrity of a host's data while eliminating unwanted traffic (malicious or not) between hosts,
that can hinder a network's performance. 

% how can this be achieved through the z-c model?
A common method of providing this guarantees is to construct modular policies using {\em network slices} \cite{gutz2012}.
A network slice is a piece of the network that can be programmed independently from the rest of the network.
If we restrict traffic to a network slice, the behaviour of the entire security policy is precisely that of that slice alone.
So, regardless of how complex a network's policy may be, we need consider only the policy of the slice corresponding
to traffic flow between two zones, to ensure proper traffic isolation between them.

%TODO: incorporate the formal notation of conduit-policy in describing this?
In our revised Zone-Conduit model, a conduit provides this network-slice behaviour.
A conduit policy does not modify packets intended for another, so to ensure traffic isolation between two zones,
we simply need to consider the policy of the conduit connecting those zones.
For example, in the policy below, traffic flow from \verb|Z1| to \verb|Z2| is strictly controlled by the policy of conduit \verb|(Z1,Z2)|.

{\footnotesize
\begin{verbatim}
Policy Company_policy { Z1 -> Z2 : https, dns; 
                        Z2 -> Z3 : http, ftp, dns;}
\end{verbatim}}

So, users from multiple policy subdomains (\eg corporate admins, SCADA engineers)
can simply manage the policies of the conduits that corresponds to their subdomain, and have guarantee of traffic isolation.
\section{System overview}
\label{sec:overview}
\begin{figure}[t]
%\center
\captionsetup{aboveskip=8pt}
\centerline{\includegraphics[width=\columnwidth]{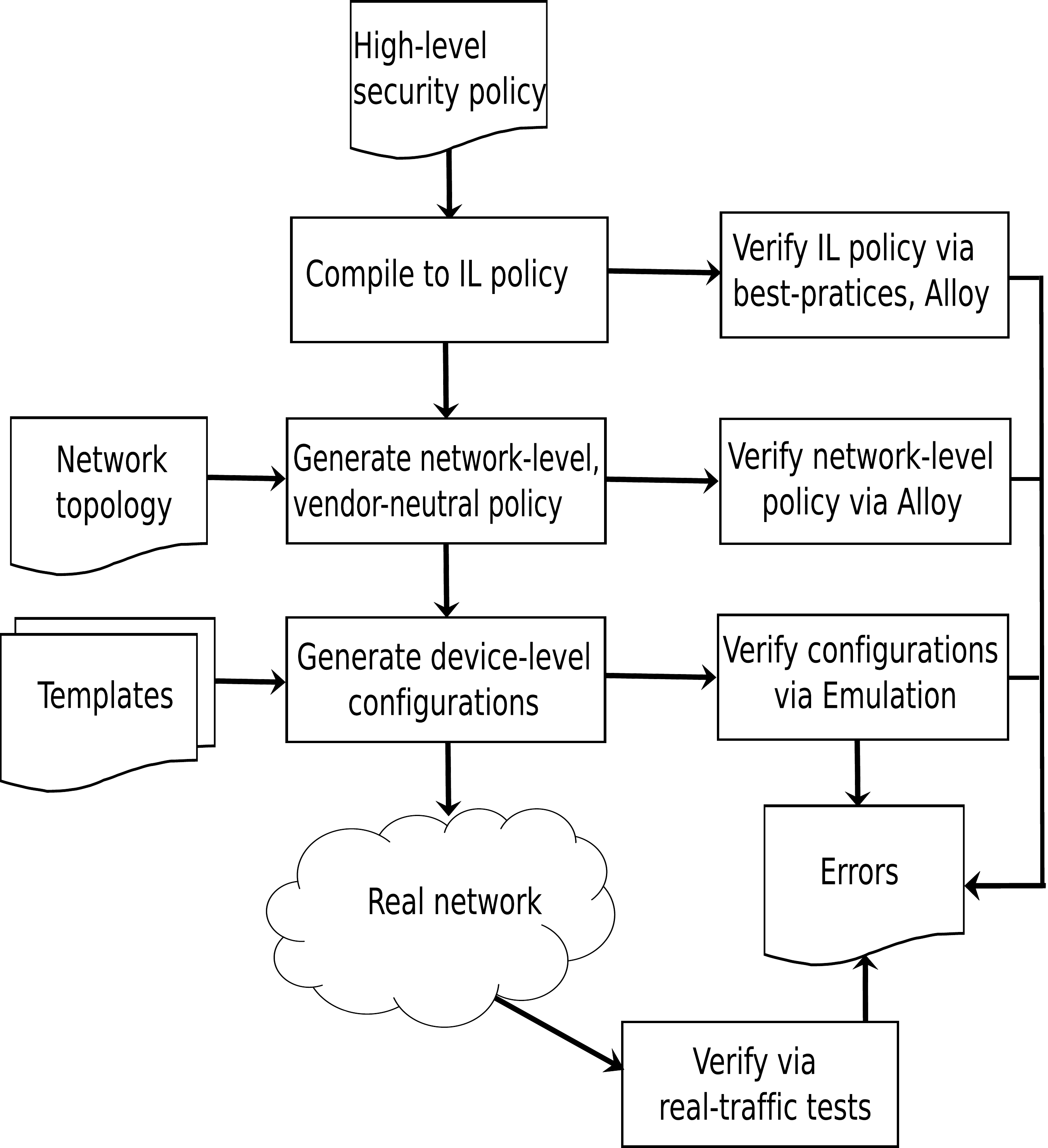}}
\caption{Firewall auto-configuration process.}
\label{fig:system-bd}
\end{figure}

We now describe our auto-configuration system design as depicted in \ref{fig:system-bd}, with the details outlined below:

% briefly describe the components
\smallskip
\noindent \textit{\textbf{High-level security policy}}: The topology-independent policy input file created using {\em ForestFirewalls}. See \ref{sec:framework} for details.

\smallskip
\noindent \textit{\textbf{Compile to intermediate-level (IL) policy}}: Parses the high-level policy to an intermediate format for checking.

\smallskip
\noindent
\textit{\textbf{Network topology}}: The input network topology described in the XML-based graph file format \textit{GraphML} \cite{GraphML}. The file contains 
information of all devices of the underlying network and their interconnections. The crucial aspects are the details of the topology near the firewalls.

\smallskip
\noindent
\textit{\textbf{Generate network-level, vendor-neutral policy}}: Down-compiles high-level policy to network-level, by coupling policy to the input network topology. See \ref{sec:network-policy}.

% putting it altogether: main BLOCK diagram of firewall auto-configuration process
%scale=0.29

\smallskip
\noindent
\textit{\textbf{Verify IL policy against best-practices, via Alloy}}: Formally checks an IL policy for SCADA best-practice violations and for correctness. Best-practice checks employ canonicalised policies while a mathematical and logic based tool, Alloy \cite{jackson2012}, finds anomalies within the policy. See \ref{sec:formal-verification}.

\smallskip
\noindent
\textit{\textbf{Verify network-level policy via Alloy}}: Formally checks network-level policy for correctness via Alloy \cite{jackson2012}. See \ref{sec:formal-verification}.

\smallskip
\noindent
\textit{\textbf{Device templates}}: A vendor and device specific meta-configurations repository that currently supports \ac{UML} IP-Tables, Cisco ASA5505 models, 
and is easily extensible.

\smallskip
\noindent
\textit{\textbf{Generate device-level configurations}}: The rendering of device-specific configurations for firewalls using the network-level policy and the device templates.

\smallskip
\noindent
\textit{\textbf{Verify via emulations}}: Device configurations are pushed to an emulated network for pre-deployment testing. Test scripts are auto-executed in this
network, to generate pathological traffic and validate configurations. See \ref{sec:testing}.

\smallskip
\noindent
\textit{\textbf{Real network}}: Device-specific configurations are pushed to hardware in a real network. At present this is conducted manually\footnote{Automation of pushing device configurations is more development than research.}, but we intend to automate it.

\smallskip
\noindent
\textit{\textbf{Verify via real traffic tests}}: Automated tests are created for the real-network, generating real-traffic, to verify post-deployment behaviour of firewall configurations. See \ref{sec:testing}.

\subsection{Network-level policy generation}
\label{sec:network-policy}

A high-level policy is implemented on a network by coupling the policy to the network topology instance. 
The resultant network-level ACL rules are vendor/device neutral.
This generic format allows easy checking of rules for inconsistencies. 
The policy generation steps are outlined below.

\subsubsection{Zone-Conduit model construction}
\label{sec:zone-cond-gen}

The system first builds a \textit{Zone-Firewall model}, containing the disjoint zones and their 
firewall interconnections \cite{ranathunga2015} using the network topology. 
Additional Firewall-Zones, Abstract-Zones and Carrier-Zones are added to the model as required. 

Next, the network conduits are defined to create the Zone-Conduit model. 
%A conduit is not an atomic device, but it implements a single security policy between two zones.
The Zone-Conduit model of the input network may not \textit{always} match that perceived by the policy creator. 
So, we must cross check the real model against that provided through the specification.
If {\em mismatched}, an error is reported indicating incompatibility.

\subsubsection {Network coupling and rule translation}

An implicit mapping between a zone and its host/subnet composition is created when defining the zones in the input network.
This mapping readily translates the high-level policy to the underlying network. 
The source and destination zone of each high-level rule can be substituted with the corresponding IP address ranges from this mapping. 
The equivalent network ACL rules can be obtained from the cross product of these IP address ranges and the original rule operator and service description.
% TODO: Actual translation is Canonical-policy -> Network-policy?
We can represent rule translation through a mapping $t: \Phi \rightarrow \Gamma$, where $\Phi$ is the high-level policy space and $\Gamma$ is the network-level policy space. 
 
Multicast rules may also be required for the correct operation of certain protocols. For instance, when OSPF is specified as a dynamic routing 
protocol by the user, multicast rules are required to enable neighbour relationships to correctly form within a single OSPF area. Similarly, 
stateful protocols (\eg TCP) require return path rules in addition to the forward path rules for correct operation. {\em ForestFirewalls} handles 
these requirements automatically, generating and incorporating any supplementary rules as necessary. 

\subsubsection {Path selection and conduit configuration}

The system identifies possible Zone-Conduit communication paths for each high-level policy rule. 
Paths that are deemed impractical are eliminated. For instance, (i) traffic cannot transit a Firewall-Zone. Firewall-Zones only enable traffic flow to and from the 
firewall but cannot forward traffic, (ii) a traffic path cannot form loops around firewalls. If a path requires a traffic packet to traverse a particular firewall interface more than 
once, it is discarded, (iii) traffic originating from or terminating at a Firewall-Zone must have a valid external path through the network.
 
Our system configures all conduits, using valid paths. A conduit implements a default \verb|deny-all| policy between its interconnecting 
zones. This strategy promotes \textit{defence in depth} \cite{isa2007, byres2005},
which prevents a single point of failure from triggering cascading security breaches across the network.

The system evaluates each conduit's firewall-interface layout to determine how the \ac{ACL} rules should be applied on these interfaces (inbound or outbound).

Our high-level policy is easily adapted to incorporate new zone additions to a network. The updated policy is swiftly re-implemented
on the network to protect the new zones.

\subsection {Formal policy verification}
\label{sec:formal-verification}

Policy-rule overlaps can cause unintended consequences. These overlaps can be \textit{redundancies} or \textit{conflicts} \cite{liu2008,yuan2006}. 
Redundant rules can be removed without affecting the semantics of a policy. 
Such rules reflect configuration inefficiencies. 
A \textit{conflict} occurs when a rule overlaps with preceding rules but specifies a different action, creating ambiguity.
Correct ordering of rules is typically required to avoid rule conflicts.
 
Our system only supports \textit{positive permissions}. 
So, we remove conflicts by design, rendering rule order irrelevant. 
Redundancies are still possible, so we need to check for these.
%TODO: include formal definition of a redundant rule

We find redundancies using a model checker: Alloy \cite{jackson2012}.
Formal model-checking is generally complex, so Alloy aims to find counter-examples to illustrate problems. 
Essentially it's a refuter \cite{jackson2012} not a prover. But, its ability to comprehensively analyse a model, 
even within finite bounds, makes it very useful \cite{jackson2012}. For Alloy based verification, see \ref{sec:verification}.

A policy also needs to be checked for SCADA best practice compliance. 
Checking policies by exhaustive comparison would be
highly inefficient.  A more efficient approach would be to derive a
unique, {\em canonical}, representation of each policy. Policy
canonicalisation can be represented through a mapping $c: \Phi
\rightarrow \Theta$, where $\Theta$ is the canonical space of
policies. Inclusion and equivalence checks on the canonical policies, help identify any violations.
For details, see \ref{sec:verification}.

%We also need to verify a policy is correct, both at a high-level and a network-level. 
%At a high-level, the rules specified using flows may contain inconsistencies. At a network-level, 
%the generated ACL rules can include redundancies. 

\subsection{Automated testing} %Pre- and post-deployment
\label{sec:testing}

% use of Netkit and Autonetkit
We use a network emulator -- Netkit -- for our pre-deploy-ment tests. The emulator is open source and enables virtual devices and interconnections using 
\ac{UML} \cite{pizzonia2008}. AutoNetkit is a tool designed to automate emulated network experimentation via Netkit 
\cite{knight2013}. We have extended AutoNetkit, with basic firewall capabilities, to generate our emulations.

When the emulated network is run, automated tests specific to the input policy create pathological traffic to verify expected firewall configuration behaviour. 
{\em ForestFirewalls} uses \textit{Expect} -- a UNIX scripting and testing utility -- to generate these test-scripts. Expect enables automated 
interactions with programs that expose a text terminal interface \cite{libes1995}. Netkit automatically launches these scripts within its 
Virtual Machines (VMs), once the VMs are running. The scripts run sequentially, with independent outcomes.
%(\ie no cascading failures, so their ordering is irrelevant). %failure of one does not affect another

Expect test scripts verify that the permits rules in a policy works correctly (\ie positive vetting).
A permit rule fails if its observed behaviour is different from that expected.
We can use a {\em result function} $R$, to track what rules fail.
For a permit rule $q^{a}_{m}$ with accept packet set  $A = \{s \in {\cal A} \mid s \prec m \}$
and a test-packet sequence $s_1 \in A$ the result function is
\begin{equation}
  \label{eq:result_f}
  R(q^{a}_{m},s_1)= 
  \begin{cases}
    0,    & \text{if $s_1$ fails for permit rule $q^{a}_{m}$}\\
    1, & \text{if $s_1$ succeeds for permit rule $q^{a}_{m}$}.
  \end{cases}
\end{equation} 

A failed permit rule means its corresponding test packets are not delivered to the intended destination.

In addition to positive vetting, we need to check that all other services not 
explicitly enabled are blocked. This negative vetting is conducted using automated, exhaustive port-scans employing \textit{nmap} and \textit{tshark}.

% real-network tests
The same test-suite can be used in the real network, post configuration deployment. The tests now generate 
live-traffic, verifying expected real-firewall behaviour.

%%%%%%%%%%%%%%%%%%%%%%% Section 6 %%%%%%%%%%%%%%%%%%%%%%%%%%%%%%%%%%
\section{Policy specification framework}
\label{sec:framework}

\begin{figure}[h]
%\center
\captionsetup{aboveskip=8pt}
\centerline{\includegraphics[width=\columnwidth]{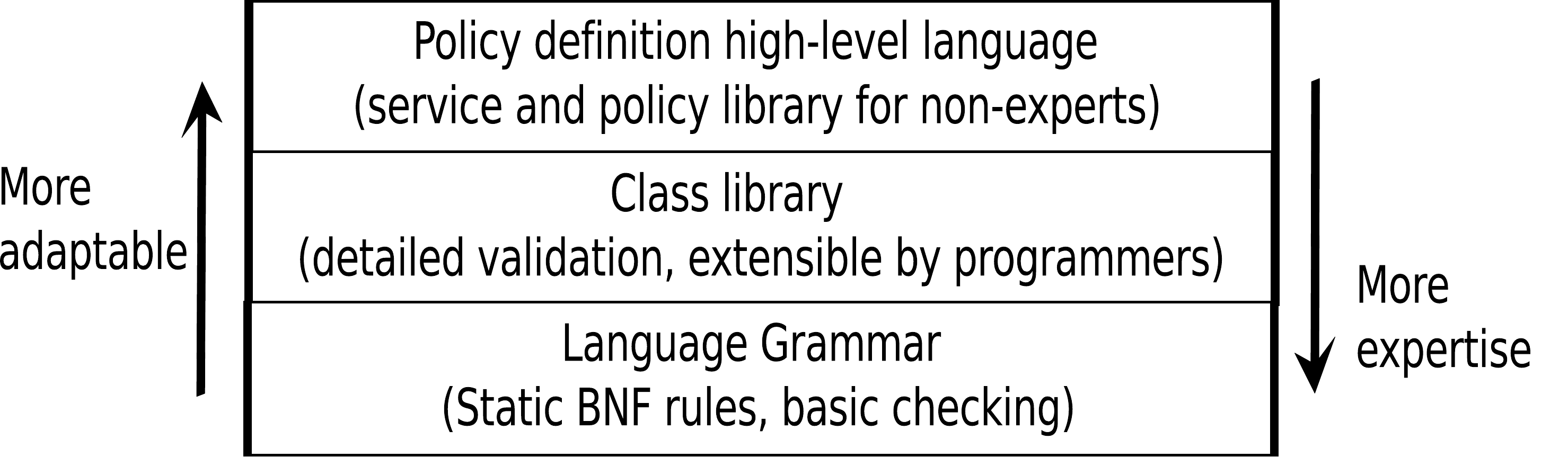}}
\caption{Policy specification in layers.}
\label{fig:layered-policy}
\end{figure}

% layered design - network/vendor indep policy to network/device centric firewall configs with verification at each layer
A useful network policy specification framework should cater for management-level policy makers as well as competent programmers. 
Policy makers need to define high-level policies to meet business goals. Programmers may wish to extend the framework to 
add more features. A layered approach (\ref{fig:layered-policy}) supports both cases. 

\subsection{A layered approach}

\smallskip
\noindent \textit{\textbf{Policy definition high-level language}}: designed primarily for non-expert users to define services and security policies, it
uses a library of services and security policies in conjunction with a simple language. The service library consists of \ac{IANA} well-known services and the policy library contains common \ac{SCADA} security policies, all easily extensible by a non-expert user. The language syntax and semantics are also intuitively simple for non-expert users. Informative warnings and errors are returned for fast debugging. See~\ref{sec:forest}. 

\smallskip
\noindent
\textit{\textbf{Class library layer}}: dedicated to expert Programmers, this layer features an \ac{OOP} based, well-defined object hierarchy that consists of 
rules for constructing protocols (\eg TCP, UDP) and services. Detailed checking of object specific attributes (\eg TCP/UDP port 
numbers are between 0-65535) are handled by their respective classes. A direct mapping between the grammar rules and the Classes makes 
the library easily extensible, but it is only intended that expert protocol engineers would extend this. Most operators would use the higher layer.

\smallskip
\noindent
\textit{\textbf{High-level language grammar}}: dedicated to the language designers, this layer consists of \ac{BNF} rules that control the 
language semantics. The grammar includes basic checking (\eg argument length, null checks), but delegates detailed checking to the 
class library layer. The rules are static and can only be altered by the language designers (us). This preserves the original objectives of 
a high-level abstraction that is intended to change slowly.

\smallskip
\noindent Our layered policy-specification architecture leads to a vendor and device neutral policy-specification framework. The system suits naive users, but the framework is easily extensible to cater for new network applications and protocols.

\subsection{ForestFirewalls high-level language}
\label{sec:forest}

Simply put, the {\em ForestFirewalls} specification language allows a user to instantiate a high-level firewall policy. Below is the definition 
(a complete example can be found in \ref{sec:concrete-example}).

{\em ForestFirewalls}' parser is currently implemented in Python and Ply (a \texttt{lex} and \texttt{yacc} implementation for Python). It 
translates a {\em ForestFirewalls} specification (\ie a \verb|.policyml| file) into its \ac{IL} 
representation using object definitions from the underlying class 
library, also implemented in Python.

\subsubsection{Service and Service-group description}

A service  is defined using the syntax

{\small
\begin{verbatim}
service <service-name> {
                protocol=<protocol-base>;
             <protocol-attributes-list>; }
\end{verbatim}}

For example, a custom implementation of HTTP, based on the above service description format is given by

{\small 
\begin{verbatim}
service custom_http {protocol=tcp;
                     tcp.dest_port=8080;
                     comment=``Internal Web'';}
\end{verbatim}}

All unspecified attribute values have defaults assigned (\eg here \verb|tcp.source_port=0-65535|).
Service specific comments are enabled via the \verb comment  field. This type of code documentation allows commentary in the 
lower tiers to be auto-generated. The aim is to help document network and device level firewall rules to avoid the common problem that rules 
cannot be deleted because no one remembers why they exist.

{\em ForestFirewalls} prohibits the description of \textit{generic services} such as \textit{all-TCP} or \textit{all-IP} for several reasons. For one, SCADA case 
studies \cite{ranathunga2015} reveal that users exploit generic rules where possible for convenience, such as allowing \textit{all-IP} 
traffic just to enable EIGRP traffic. However, far more services than necessary are thus admitted through firewalls. 

Secondly, such inherently 
broad services don't contribute towards forming well-defined security policies. They cloud the ability to accurately see the type of 
cyber threats a network is being protected from. 

A {\em service} can be formally represented by tuple $(Pr,PrA,$ $PrV)$ where $Pr \in \Lambda; \Lambda=\{all \ IP \ protocol \ numbers\}$ and $PrA \subset \Upsilon, PrV \subset \Pi;$ $\Upsilon=\{tcp/udp/icmp \ etc. \ protocol\ attributes\}$ and $\Pi=\{protocol \ attribute \ values \}$.

We use a \verb|service-group| to bundle services and other service-groups (\ie nesting is allowed)

{\small
\begin{verbatim}
service_group <group-name>{
                      <service-or-group-list>}
\end{verbatim}}

Service-groups provide a level of indirection, so application protocols used to achieve network functionality (\eg Web services) 
can change without needing policy alterations.
%They also provide convenience but are explicit. 

A \verb|service-group| is really a set $SG=\{(Pr_i,PrA_i,PrV_i)$ $| Pr_i \in \Lambda, PrA_i \subset \Upsilon\, PrV_i \subset \Pi\}$.
The specification supports set operations: union (\verb|,|), intersection (\verb|^|) and difference (\verb|\|), 
so, new service groups can be constructed by applying these operators on existing service groups.

The following snippet defines a \verb|service-group| containing various example Web services:

{\small
\begin{verbatim}
service_group Web { http, https, dns }
\end{verbatim}}

\subsubsection{Zone-group description}

A \verb|zone-group| bundles a set of zones or other zone-groups and  is defined using the syntax

{\small
\begin{verbatim}
zone_group <group-name> {<zone-or-group-list>}
\end{verbatim}}

A \verb|zone-group| is a as a set of disjoint zones $ZG \subset \Omega$ where $\Omega=\{all \ zones \}$.
Multiple zone-group declarations are checked for duplicates to minimise code redundancy.

The snippet below describes an example zone-group, depicting \verb three_zones : a set of zones in a network which is made up of 3 disjoint zones.

{\small
\begin{verbatim}
zone_group three_zones {corp, scada, dmz}
\end{verbatim}}

We also allow similar syntax and set operators for defining groups of TCP/UDP ports or ICMP types.

\subsubsection{Policy-rule description}

A high-level policy rule can be defined as below with an \verb|operator| to direct the {\em inter-zone flow} explicitly. 
The end-zones defined by \verb|zone-or-group-name| have traffic flow of type \verb|service-or-group-name| enabled.

{\small
\begin{verbatim}
policy_rule <rule-name> {
            <first-zone-or-group-name> operator 
            <second-zone-or-group-name> : 
            <service-or-group-name>  } 
\end{verbatim}}

\indent \indent where \verb|operator == `->' or `<->'| \\

For example, the following policy rule models the capabilities of a \verb Corp_  \verb orate_zone  with regards to \verb Web  traffic

{\small
\begin{verbatim}
policy_rule corp_web_rule {
                 Corp_zone -> DMZ : Web}
\end{verbatim}}

The above description can be represented as a policy rule $q^{a}_{m}$ with accept packet set  $A = \{s \in {\cal A} \mid s \prec m \}$ where predicate $m=(s.hdr.source\_address \ in \ Corp\_zone \ \wedge \ s.hdr.dest\_$ $address \ in \ DMZ \ \wedge \ s.hdr.service==Web)$.

\subsubsection{Policy description}

A \verb|rule_group| object is used to hold one or more security policy rules and can be defined using the following format

{\small
\begin{verbatim}
rule_group security_policy {
                     <rule1>, <rule2>, ... }
\end{verbatim}}

A security policy is a set of rules $RS=\{q^{a}_{m_i} ; m_i-predicate \}$, each rule has an accept packet set  $A = \{s \in {\cal A} \mid s \prec m_i \}$.

Similarly, we can specify firewall reporting policy using a \verb|reporting| \verb|_rule| object \cite{ranathunga2015P}

{\small
\begin{verbatim}
reporting_rule reporting_policy {
                            <attributes-list>}
\end{verbatim}}

A global \verb|policy| object then encapsulates security policy with reporting policy \cite{ranathunga2015P}

{\small
\begin{verbatim}
policy <policy-name> { security_policy; 
                       reporting_policy; }
\end{verbatim}}

\subsubsection{ForestFirewalls library file imports}

%- why library file import is critical
%- our simple yet effective architecture: pros
%- example usage of library files?
In large industrial control plants, multiple policy sub-domains exist (\eg corporate admins, SCADA engineers, network engineers, different departments) that set their own policies to be applied to the network components they own or manage. Namespace importation facilitates such distributed policy management and promotes reuse of policy. 
For instance, we can imagine the ISA creating a best practice ruleset for SCADA as a baseline for new installations.   

\newlength{\figurewidthA}
\setlength{\figurewidthA}{0.31\textwidth} 
\begin{figure*}[ht]
  \captionsetup{aboveskip=8pt}
  \centering
  \subfigure[Four rules indicated by (overlapping) rectangles.] % caption for subfigure a
  {
    \includegraphics[width=\figurewidthA]{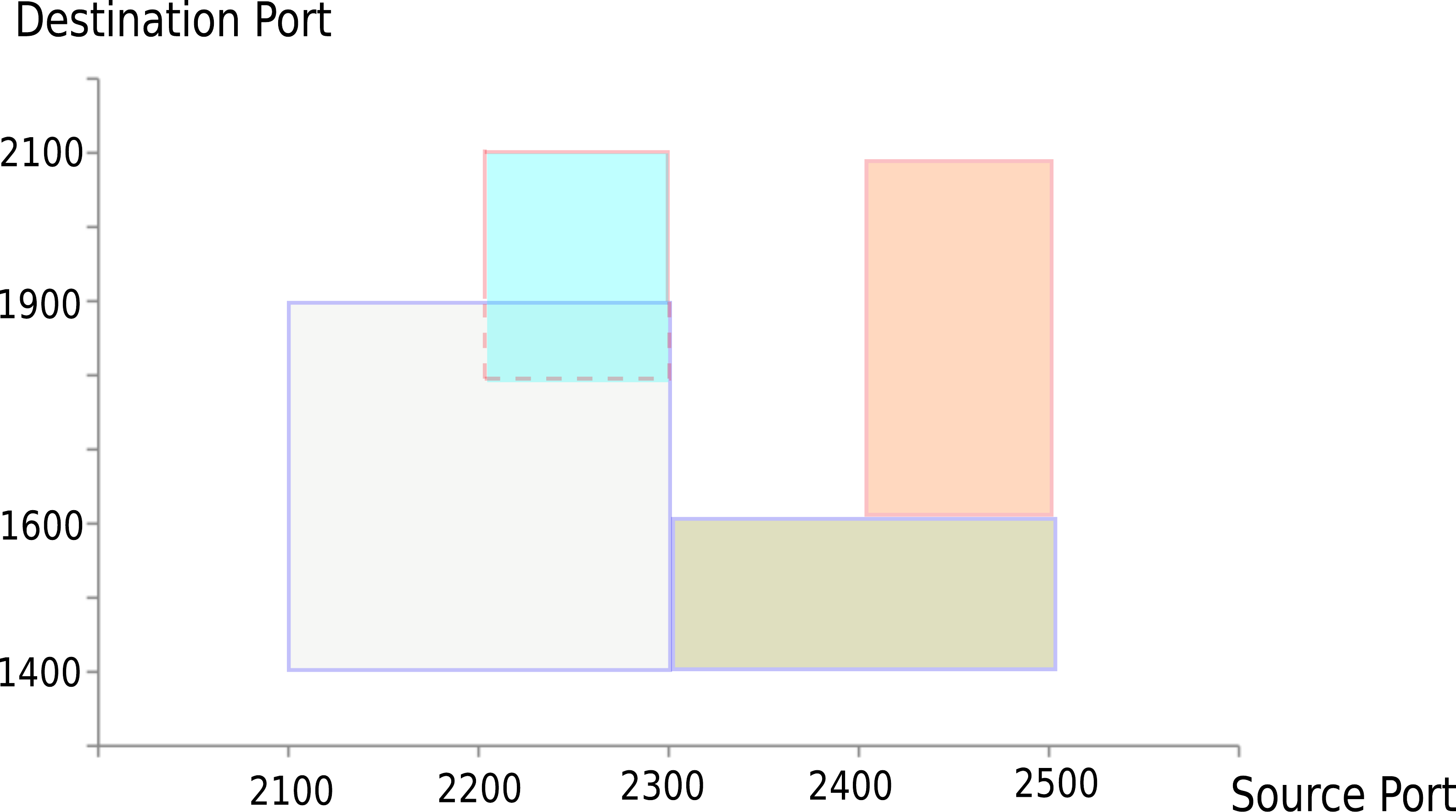}
    \label{fig:polygon1}
  }
  \hfill
  \subfigure[Five rules producing an equivalent policy to (a).] % caption for subfigure b
  {
    \includegraphics[width=\figurewidthA]{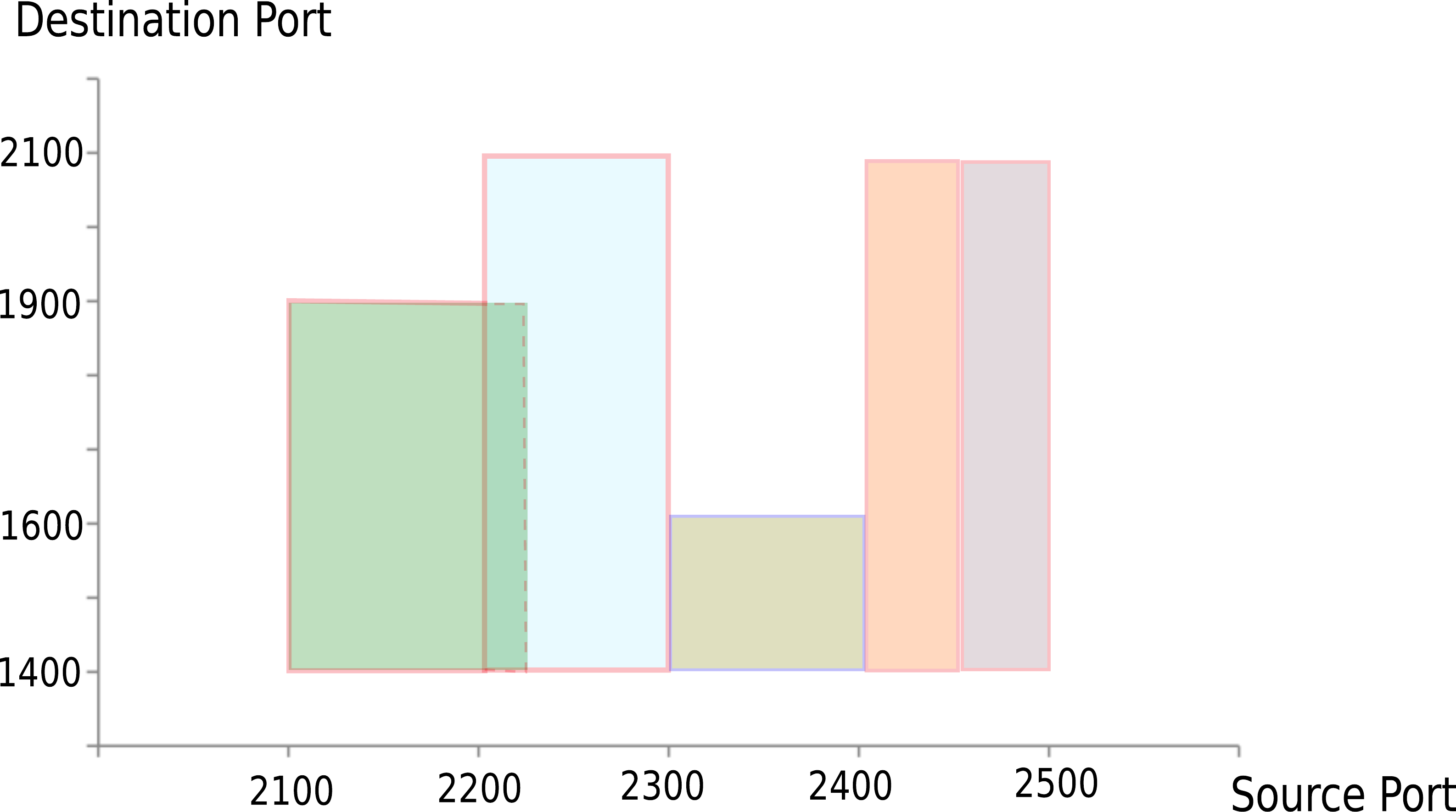}
    \label{fig:polygon2}
  }
  \hfill
  \subfigure[Horizontal partitions of polygon in (a) or (b).] % caption for subfigure b
  {
    \includegraphics[width=\figurewidthA]{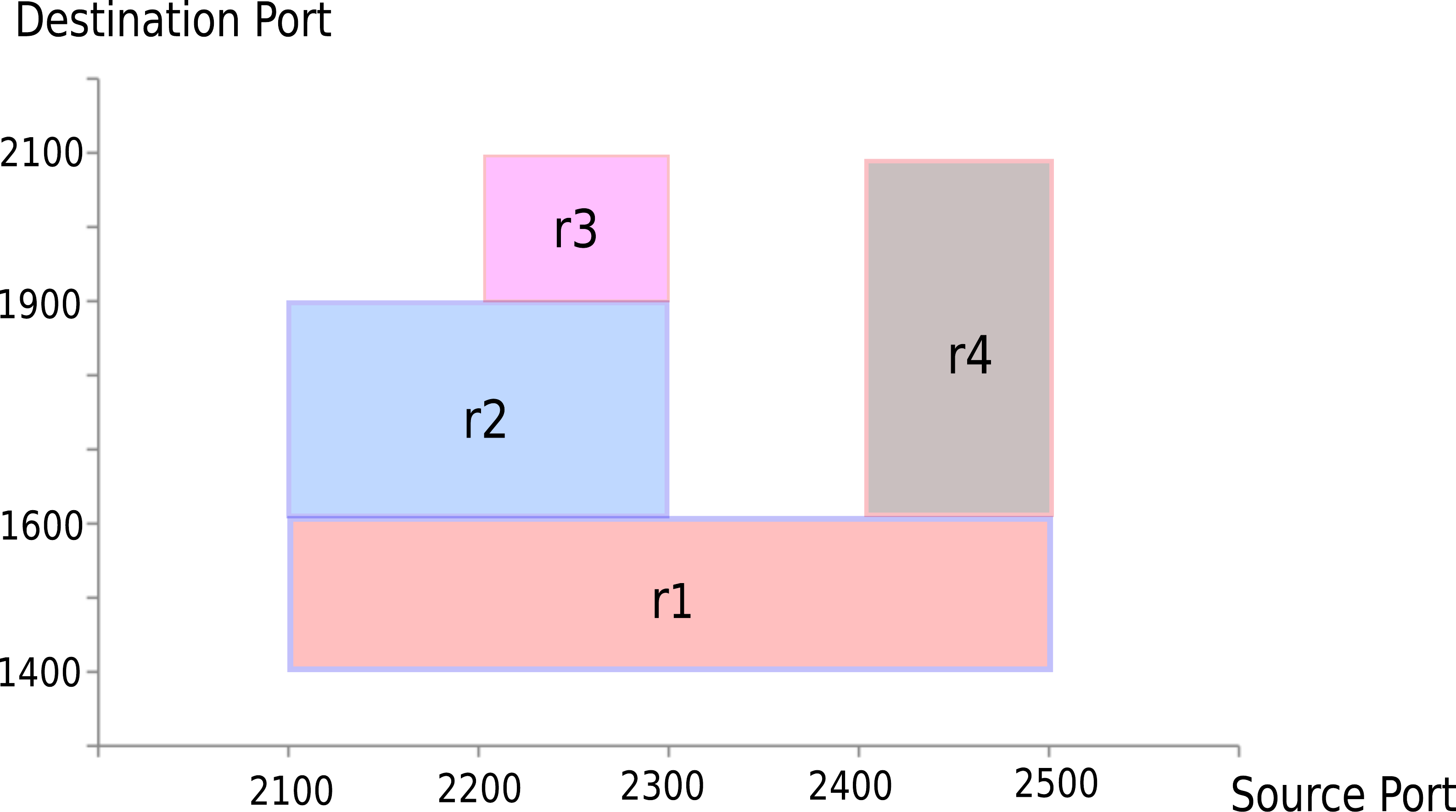}
    \label{fig:polygon3}
  }
  \caption{Canonicalisation of distinct rule sets of the same policy.}
  \label{fig:policy1}
\end{figure*}

Non-expert programmers may, however, be unfamiliar with the use of a complex namespace library with many features. We developed a namespace hierarchy 
that is simple, yet provides rich features for managing and reusing namespaces. {\em ForestFirewalls'} namespace hierarchy consists of 
generic library definitions for all users with additional support for custom namespace creation. 
This allows to compose distributed policies into a coherent policy set, free of inconsistencies.
%This custom namespace feature allows to compose distributed policies into a coherent policy set, free of inconsistencies.

%%%%%%%%%%%%%%%%%%%%%%% Section 7 %%%%%%%%%%%%%%%%%%%%%%%%%%%%%%%%%%
\vspace{-2mm}
\section{Verification} 
\label{sec:verification}

Once a high-level policy is parsed by {\em ForestFirewalls}, it is stored in IL code and needs to be checked for \ac{SCADA} 
best-practice violations as well as for correctness.

\vspace{-2mm}

\subsection{Best-practice compliance}

%TODO: refer to the policy comparison paper for further details?
Two policies with different rule sets, can have the same underlying
semantics (\ie they allow the same set of services between
zones). \ref{fig:polygon1} and
\ref{fig:polygon2} illustrate the idea based on TCP port
filtering of single packets. Each rectangle indicates the allowed
packets of a single rule. Combined, the rules cover the same set of
allowed packets. 

Comparing policies exhaustively
is highly inefficient. So, we derive a unique, {\em canonical}, representation of each policy
to efficiently compare them. In our canonicalisation mapping
$c: \Phi \rightarrow \Theta$, where $\Theta$ is the canonical space of
policies, all equivalent policies of $\Phi$ map to a
singleton. For $p^X, p^Y \in \Phi$, we note the following (the proof 
follows the definition)
\begin{lem}
  Policies $p^X \equiv p^Y$  iff $c(p^X)=c(p^Y)$.
\end{lem}

Thus, comparison is eased by the canonicalisation of policies.
We illustrate the idea using TCP policy rules (\ref{fig:policy1})
and dissect the polygon formed in our example policy into
horizontal partitions (Figure \ref{fig:polygon3}), using a Rectilinear-Polygon
to Rectangle conversion algorithm \cite{gourley1983}. Each partition is
chosen to provably guarantee its uniqueness. Canonical policy
elements are derived by translating each partition back to a
rule and ordering the resulting rule-set uniquely in increasing
IP protocol number and source and destination port numbers.
We find a unique partition quickly rather than a guaranteed
minimal partition. The result is a deterministic, ordered set of
non-overlapping rules.

So, intra-policy verification can be performed  by comparing canonical policy components.  For instance

\smallskip
\indent \indent Is $c ( p^{Z1 \to Z2} ) = c ( p^{SCADA \to Corp} )$ ?

\smallskip
Another useful notation, linked to the goal of policy comparison,
is that policy $P^A$ {\em includes} policy $P^B$. 
Particularly in SCADA networks, the notation helps evaluate whether the policies are
compliant with industry-recommended practices in \cite{stouffer2008,byres2005}. These guidelines 
specify potentially dangerous services (\eg HTTP) that should be {\em prohibited} from traversing inbound and/or outbound from the
(protected) SCADA-Zone. A violation here, means increasing the vulnerability of a SCADA-Zone to cyber attack.
Unlike corporate networks, this increased exposure could potentially render SCADA systems unavailable, 
and cause significant financial loss or at worse loss of human lives.

%We would like, for instance, to evaluate whether our policies are compliant with industry-recommended guidelines.

We indicate that a policy {\em complies} with another if it is more restrictive and define the following
\begin{defn}[Inclusion]
  A policy $p^X$ is {\em included} in $p^Y$ on ${\cal A}$ iff $p^X(s)
  \in \{p^Y(s), \phi\}$, \ie $X$ either has the same effect as $Y$ on
  $s$, or denies $s$, for all $s \in{\cal A}$. We denote inclusion by
  $p^X \subset p^Y$.
  \label{def:includes}
\end{defn}

So, we can now evaluate whether an input policy adheres to the SCADA best practice policy using an inclusion check

\smallskip
\indent \indent Is $p^{Input} \subset p^{BestPractice}$ ?

\subsection{Policy correctness}
A partial snippet of the Alloy language specification (\ie \verb|.als|) file auto-generated by {\em ForestFirewalls} for the IL policy is shown in  Figure 8 in the Appendix.
It depicts a formal model with 3 signatures:
\verb|Service|, \verb|PolicyRule| and \verb|SecurityPolicy|. In our initial model, a 
\verb|Service| has the basic members: \verb|ip_protocol|, \verb|source_port|, \verb|dest_port| and \verb|icmp_type|. Of these 
members, only \verb|ip_protocol| is mandatory.
% (the multiplicity keyword \verb|some| requires at least one element). 

A \verb|PolicyRule| has 4 members: \verb zone1  and \verb zone2  to capture the zone names, an \verb operator  and a single 
\verb service  element. 
%The \verb operator  is a set of integers containing \textit{\{1\}} (representing a uni-directional permit: \verb|->|) or \textit{\{1,\,2\}} (representing a bi-directional permit: \verb|<->|). The bi-directional permit is deconstructed as \verb|->| \textit{\{1\}} and \verb|<-| \textit{\{2\}} for simplicity, but the latter operator is omitted in the high-level 
%syntax. 
The global constraints are partially shown (lines 17--27 in Figure 8), requiring the universal set (\textit{Univ}) of \verb|PolicyRule| to
comprise entirely of rules in the policy. \textit{Univ} of \verb|Service| must also comprise of \verb|PolicyRule| services. 
 
Predicates can determine if two given rules or services overlap (not shown). \verb|Service| overlaps are found by
computing their intersection and testing if the result has members. \textit{String} type members (\eg \verb|zone1|, \verb|zone2|) can be directly compared. 
\verb|PolicyRule| overlaps are checked similarly.
 
%TODO: may be discuss about how scope is determined: #signatures x 3 (rule of thumb as per Alloy text)?
A `no\_rule\_overlaps' assertion (also not shown) is used to locate distinct rules with overlapping criteria. 
If found, a counter-example is returned, indicating potential inconsistencies in the high-level policy. 
Counter-examples can be inspected through Alloy's GUI to find the underlying cause(s).

Currently, Alloy is itself run manually and output counter-examples help debug data. We do not auto-correct rules as this requires human discretion.

When overlaps are absent in the high-level policy, there should also be none in network-level policy,  in theory. But we cannot simply `trust' our system to always correctly generate network-level policy.
So, we re-check the generated policy for overlaps and verify the fact.

The Alloy export generated for network policy verification is similar to the high-level export. The key exception is the source and destination 
zone names are now replaced with IP address ranges in an \verb|ACLRule|. Additionally, the \verb|Service| signature also has members 
depicting protocol state. We also define an assertion here to check for \verb|ACLRule| overlaps.

%%%%%%%%%%%%%%%%%%%%%%% Section 8 %%%%%%%%%%%%%%%%%%%%%%%%%%%%%%%%%%
\section{A concrete example}
\label{sec:concrete-example}

\begin{figure}[ht]
%\center
\centerline{\includegraphics[width=\columnwidth]{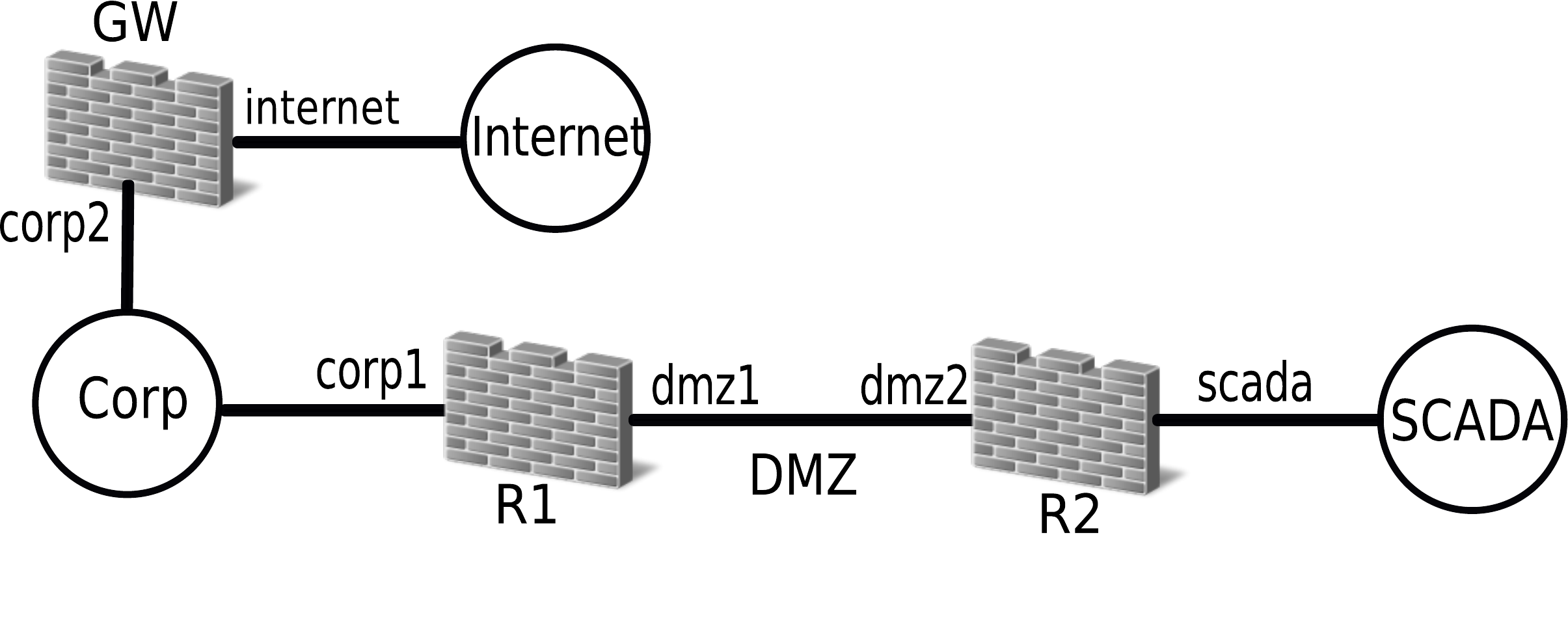}}
%\hfill
\caption{The \ac{SCADA} network under study. \texttt{Corp} and \texttt{SCADA} are the 
corporate and SCADA subnets while the firewalls are R1, R2 and GW.}
\label{fig:scada-net}
\end{figure}

We show here a concrete example, illustrating our methodology and the prototype system. The example is based on an actual SCADA case study \cite{ranathunga2015} with the multi-firewall network configuration shown in \ref{fig:scada-net}. Due to security concerns and non-disclosure agreements, a modified version of the real network is presented for discussion. Steps have been taken to keep the core security strategies intact, but details such as IP addresses are anonymised.

% describe the SUC: subnets, firewalls, nature of hosts in the subnets and type of services permitted across firewalls
\indent R1 is a Cisco ASA 5505 firewall and R2 and GW are Linux IPtables firewalls.
\ref{fig:scada-net} shows these best already. The subnet summary is below.
% R2 points to \verb|SCADA| and \verb|LAN| while GW points to \verb|Corp| and the Internet. 
% with 2 active physical interfaces pointing to \verb|Corp| and \verb|LAN|.

\smallskip
\noindent
\textit{\textbf{The Corporate network (Corp):}} Provides access to business applications and the Internet. 

\smallskip
\noindent
\textit{\textbf{Demilitarised Zone (DMZ):}} Responsible for enabling connectivity between R1 and R2. 
The distinct vendor firewalls provide \textit{defence in depth} \cite{byres2005} by having different modes of failure and firewall-software redundancy. 

\smallskip
\noindent
\textit{\textbf{The SCADA network (SCADA):}} Responsible for providing networked access to plant equipment.

\smallskip
\noindent
\verb|Corp| and \verb|SCADA| could accommodate 2,046 and 65,534 hosts respectively. \verb|Corp| hosts management 
 workstations, a HTTP server, a HTTPS server, a FTP server, an Email server, a syslog server and a DNS server. \verb|SCADA| has
\verb|Oracle| database servers, management workstations and a HTTPS server.

\subsection{Policy goals}

We consider a simple policy which nonetheless covers many of the aspects that occur in more complex, real-life SCADA policies
\cite{ranathunga2015}. Its premise is that internal corporate users are trusted, but are restricted to use \textit{safe} protocols when accessing \verb|SCADA|. 
External users are allowed access only to content that is explicitly made public. The policy has these goals:
\begin{sitemize}
\item \verb Corp  hosts can access the \verb Oracle  servers and the HTTPS server in \verb|SCADA|. They can also access all HTTP, HTTPS, DNS resources on the Internet.
\item \verb SCADA  hosts can access Web, Email and DNS servers on \verb Corp . Additionally, they can perform file transfer using FTP and HTTP with respective \verb Corp  servers.
\item Corp's HTTP, FTP, Email servers are Internet accessible.
%\item External hosts can access the HTTP, FTP and Email servers in \verb Corp .
\item R1, R2 can be managed from \verb Corp  using HTTPS and SSH. R2 can be managed from \verb SCADA  using SSH. R1 can also be managed from R2 using SSH.
\item A Syslog server located in \verb|Corp| stores firewall logs.
\item OSPF is enabled across the entire site.
\item Firewall reporting is enabled for policy verification.
\end{sitemize}

\subsection{Implementation}
A partial snippet of the {\em ForestFirewalls} high-level description implementing the above policy goals is depicted in Figure 10 in the Appendix. 
We start by importing the required library files containing the predefined lists of \ac{IANA} well known services. 
Next, the Zone-Conduit security model is supplied as a GraphML file. The zones within this model can be grouped as necessary (lines 9-15),
to simplify the specification process and increase readability. We also define custom port-groups, services and service-groups as needed.
 
A passive mode FTP data service is enabled through the firewalls (lines 18--20) as it's the best-practice approach\cite{stouffer2008,byres2005}. 
Ping is also defined for connectivity tests. High-level rules are defined to match the policy goals listed earlier. 
Finally a policy object is used to hold all the rules (line 48).
 
\subsection{Procedure and Results}
\begin{figure}[t]
%\center
\centerline{\includegraphics[scale=0.70]{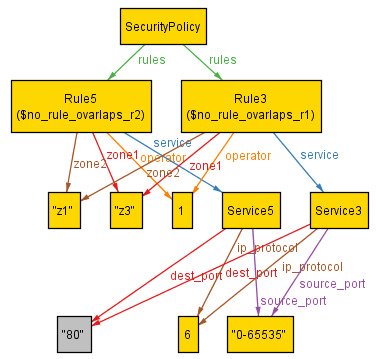}}
%\hfill
\caption{Counter-example thrown by Alloy, indicating a high-level policy error.}
\label{fig:alloy-error}
\vspace{-5mm}
\end{figure}

%TODO: may be discuss about how scope is determined: #signatures x 3 (rule of thumb as per Alloy text)?
Once the high-level policy is parsed, the corresponding Alloy export is generated by {\em ForestFirewalls} for verification.
Checking the `no\_rule\_overlaps' assertion (not shown) returns a counter-example, indicating potential inconsistencies in the specification. 
Upon inspection of the counter-example details in Alloy (\ref{fig:alloy-error}), we see that rules enabling HTTP services (\verb|ip_protocol=6|, \verb|dest_port=80|) between zones \verb|Z3| and \verb|Z1|, initiate the overlap. The root cause is the \texttt{web} and \verb|file_transfer| service-groups in the high-level policy (lines 22 and 25 in Figure 10), both containing HTTP. Once this is rectified (remove HTTP from \verb|file_transfer|), no further counter-examples are found by Alloy.

 \begin{figure*}[t!]
\captionsetup{aboveskip=8pt}
\centering
\subfigure[Zone-Firewall model.] % caption for subfigure a
{
	\includegraphics[scale=0.275]{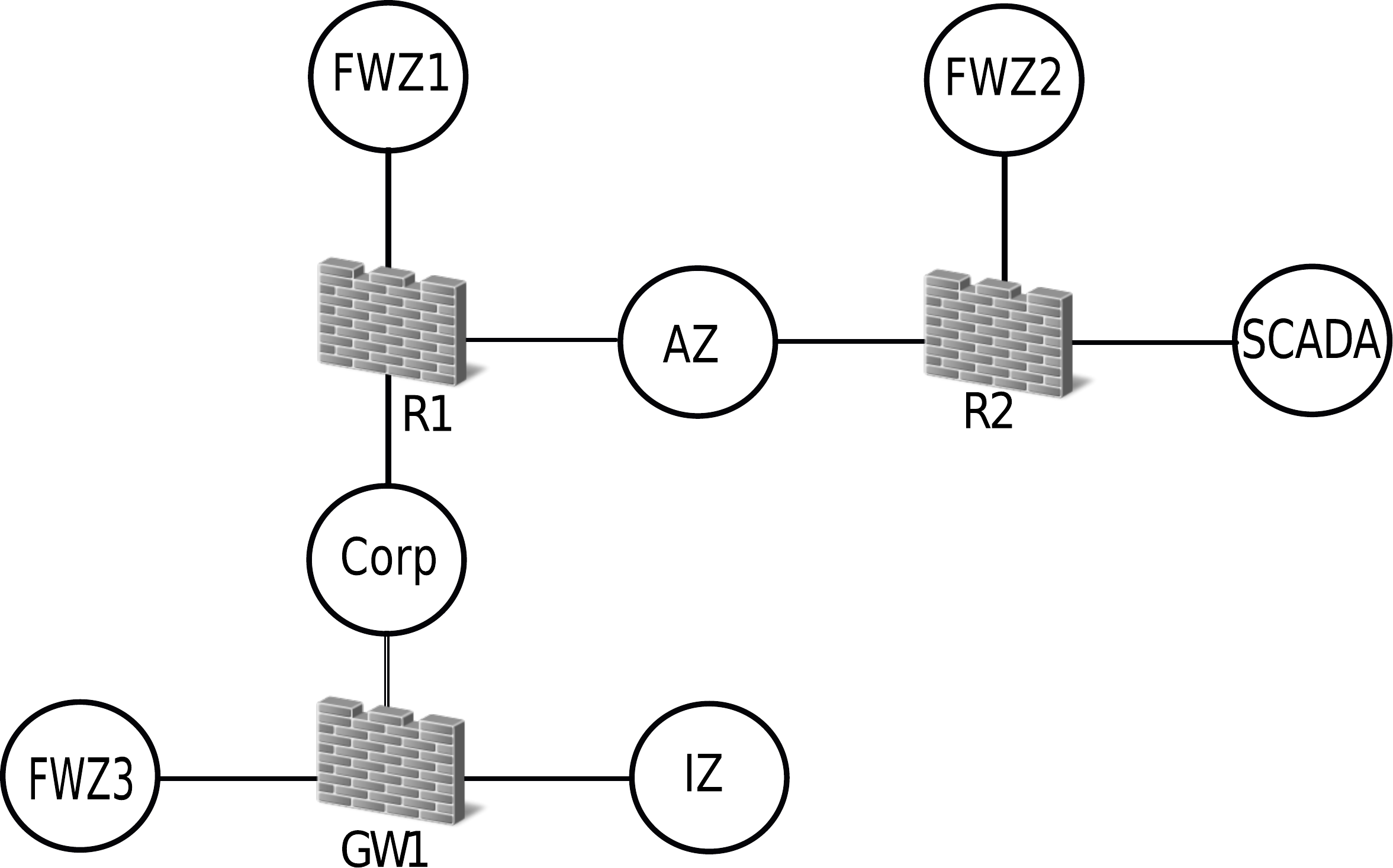}
	\label{fig:zfw2}
	% original scale=0.257
}
\hspace{0.1cm}
\subfigure[Zone-Conduit model of (a).] % caption for subfigure b
{
	\includegraphics[scale=0.275]{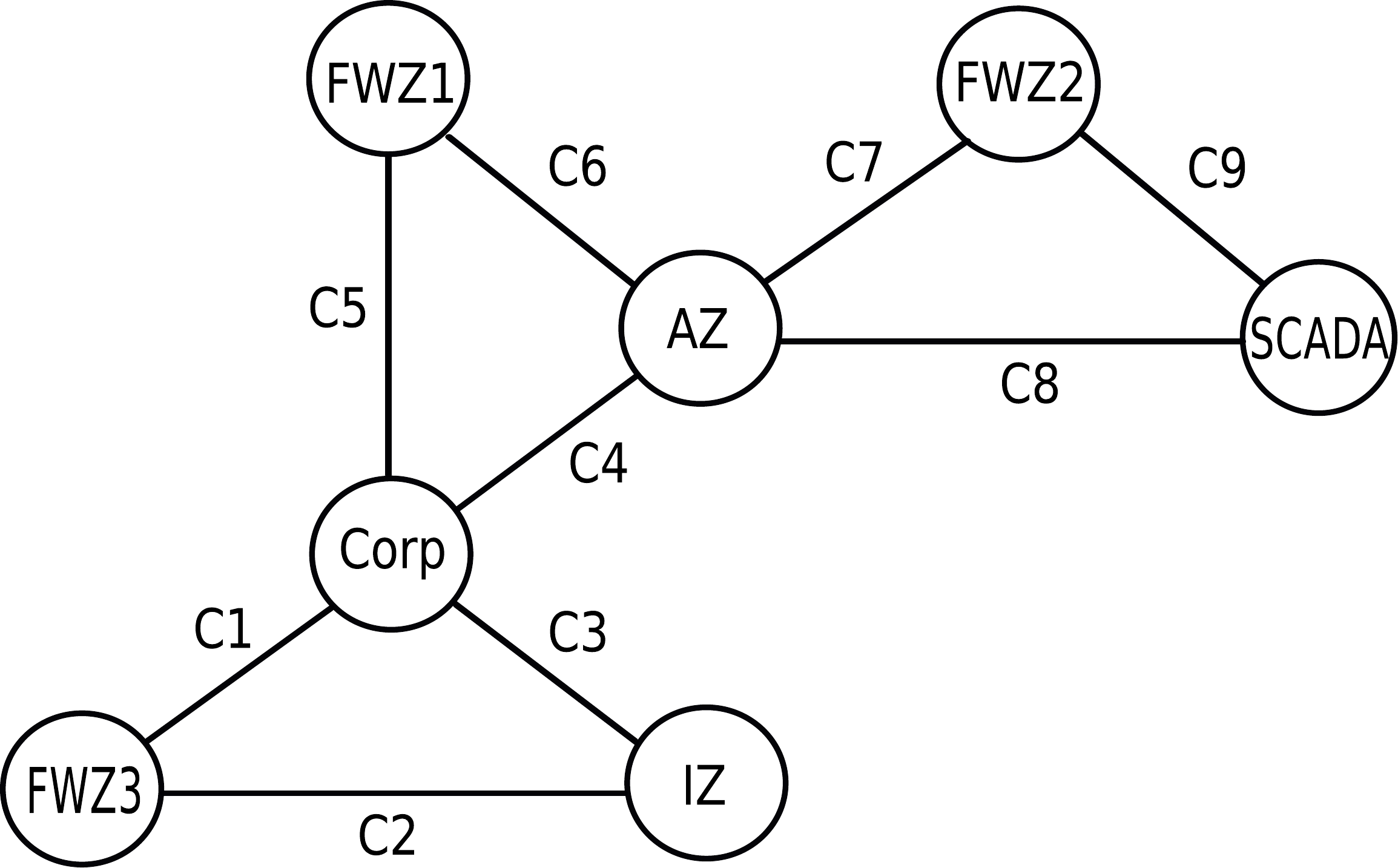}
	\label{fig:zcd2}
	% original scale=0.257
}
\caption{System generated security models of the network.}
\label{fig:sec-models2}
\end{figure*}

\begin{table*}[t!]
\caption{High-level comparison of Original vs Generated configurations (LoC - Lines of Code).} % title of Table
\centering % used for centering table
\begin{tabular}{c c c c c c} % centered columns (4 columns)
\hline\hline %inserts double horizontal lines
Type &  \multicolumn{1}{p{2cm}}{\centering Device-level \\ LoC} &  \multicolumn{1}{p{3cm}}{\centering Obsolete-ACL \\ count} &  \multicolumn{1}{p{3cm}}{\centering Generic permit- \\ rule count} &  \multicolumn{1}{p{3cm}}{\centering Intra-ACL \\ interaction count} \\ [0.5ex] % inserts
%Type & High-level LoC & Device-level LoC & Obsolete ACL count & Generic rule count & Intra-ACL Interaction count \\ [0.5ex] % inserts table
%heading
\hline % inserts single horizontal line
Original case study &  2720 & 2 & 324 &167 \\ 
ForestFirewalls generated &  714 & \textbf{0} & \textbf{0} & \textbf{0} \\
\hline %inserts single line
\end{tabular}
\label{table:comparison} % is used to refer this table in the text
\end{table*}

Figure 9 in the Appendix shows the ACL-allocation map for \verb|R1|, indicating how ACLs are assigned to the firewall's interfaces. 
 
Figure 11 in the Appendix partially shows the generated vendor-neutral ACL rules. Note the explicit \textit{deny all} rule supplementing the explicit permit rules at the end.
The step also outputs the Zone-Firewall and Zone-Conduit models of the input network as graphical output (\ref{fig:sec-models2}). 
 
The network-level Alloy exports have 828 \verb|Service|, \verb|ACLRule| objects.
Assertion checks here yield no counter-examples. 
% (\verb|Service|, \verb|ACL-| \verb|Rule|) objects. 

 %TODO: is there anything else to add about the device config generation?? 
The device-specific configurations are rendered from the network-level policy, using vendor and device specific \textit{Mako} templates.
Mako is a template library written in Python \cite{mako}, enabling fast and easy integration into {\em ForestFirewalls}.

The device-level configurations generated were first auto-deployed 
to a Netkit-based emulated network. Once the Netkit \ac{VM}s booted up, the test scripts were run automatically. 
The emulation results confirmed that the firewalls correctly admitted the services explicitly enabled through the high-level policy.
Moreover, automated exhaustive port-scans using \textit{nmap} and \textit{tshark} showed that no additional 
services were permitted through the firewalls.
 
Post emulation testing, the device configurations were deployed to the real-network. Although we aim to 
automate this deployment, it is currently done manually 
as its not seen as an error-prone step in modern configuration tools \cite{cisco2010,cisco2014}.

Once deployed, we re-executed the emulation test scripts on hosts in 
the various zones of the network. The tests confirmed that the services enabled by the input policy were passing across firewalls as expected, and  
port scans confirmed no additional services were allowed through.

\subsection{Cyber attack mitigation} 

%relate to our case studies 
% - what attacks were they vulnerable to 
% - how did our system reduce/remove such vulnerabilities 
We have studied firewall configurations from 7 real SCADA networks to date \cite{ranathunga2015} (including the network discussed), 
and found the following serious cyber-security vulnerabilities:
\begin{sitemize}
\item{{\em Insecure protocols enabled through explicit generic rules:} all-TCP, all-UDP and all-IP traffic flow were explicitly enabled inbound to SCADA, permitting far more services than necessary. Inherently unsafe protocols such as FTP and HTTP were thus allowed into SCADA. HTTP for instance, is known to transport worms and attacks. These generic rules significantly increased the vulnerability of a SCADA-Zone to cyber attack.}
\item{{\em Insecure protocols enabled through implicit rules:} incorrect use of implicit rules such as Cisco security levels \cite{cisco2010} enabled {\em all-IP} traffic to flow between Corporate and SCADA zones, making the latter more prone to cyber attack.}
\item{{\em Direct communication enabled between the SCADA-Zone and the Internet:} allowing so, clearly violated industry best-practices and significantly elevated the risk of a cyber attack on the SCADA-Zone.}
\item{{\em Insecure protocols enabled through explicit and implicit-rule interactions:} in some cases, insecure protocols (\eg HTTP) were explicitly disallowed inbound to the SCADA-Zone, but were then implicitly enabled into the same zone, exposing the latter to cyber attack.}
\end{sitemize}

{\em ForestFirewalls} addresses each of the above cyber security vulnerabilities comprehensively. 
A comparison of the firewall configurations observed in the case study discussed and those those generated by {\em ForestFirewalls}
are shown in \ref{table:comparison}. It shows that there are \textit{no explicit generic permit rules} generated by our system (\textit{i.e.,}
all-TCP, all-UDP or all-IP based rules). Eliminating these rules prohibits unwanted services from being enabled implicitly between zones.
Additionally, {\em ForestFirewalls} only utilises explicit rules to enable both firewall management and non-management traffic.
So, the use of implicit rules such as {\em Cisco security levels} are removed altogether. Doing so, also prevents interactions between explicit and implicit rules, so,
one cannot override the other to accidentally enable services.

Also shown in \ref{table:comparison}, there are \textit{no redundant ACLs} generated by {\em ForestFirewalls}. 
Each ACL serves a purpose and is assigned to an active firewall interface.
There are also \textit{no intra-ACL rule interactions} in the ACLs generated, making these configurations comparatively more efficient.

Our system also formally checks a security policy against industry best practices for compliance. 
Any violations are flagged for the user to resolve and policy compilation stops until the issues are rectified. 
The step prevents direct communication between SCADA and the Internet being enabled.

%We discussed with security consultants and distilled the core policy from the original case study.
%We also used Cisco ASA5505 and Linux IPtables devices instead of the Cisco IOS routers used originally. 
%But, our results show that there are significant improvements achieved by {\em ForestFirewalls}. 

These are almost obvious consequences of our design approach but the real firewalls \cite{ranathunga2015} had all of these defects.
Our system only requires 80 high-level LoC (only 41 LoC are policy specific) to generate 714 device-level LoC to configure all 3 firewalls in the case study discussed. 
This high-level policy with only 80 LoC has replaced 2720 attack and error prone, inefficient, device-level LoC of the original case study!
In fact for all 7 real SCADA networks we studied, it was possible to replace their total 7694 firewall-level LoC with only 271 high-level {\em ForestFirewalls} LoC.

%%%%%%%%%%%%%%%%%%%%%%% Section 9 %%%%%%%%%%%%%%%%%%%%%%%%%%%%%%%%%%
\section{Conclusions and Future Work}
The current manual approach to firewall configuration is complex and error prone. Various firewall vendor tools attempt to facilitate high-level configuration, 
but these lack flexibility in specifying detailed traffic restrictions and do not reduce the configuration burden.

{\em ForestFirewalls} greatly reduces the configuration burden, and by use of high-level abstraction, templates and graphs, offers a simple and
manageable approach to SCADA firewall configuration. Our system guarantees configuration accuracy through stage-wise validations employing SCADA best-practices, a formal verification tool (Alloy), and emulation based pre-deployment tests. The system gives users assurance of the generated device-level configurations
delivering the expected firewall behaviour prior to deployment. The ability to configure a group of firewalls at once makes {\em ForestFirewalls} scale at lower cost.
\vspace{-3mm}


%\section*{Acknowledgment}

%\begin{thebibliography}{1}

%\end{thebibliography}

\nocite{byres2005}
\nocite{cheswick2003}
\nocite{stouffer2008}
\nocite{byres2012}
\nocite{isa2007}
\nocite{jamieson2009}
\nocite{mayer2000}
\nocite{bellovin2009}
\nocite{cisco2010}
%\nocite{yed}
\nocite{ranathunga2015}
\nocite{yuan2006}
\nocite{liu2008}
\nocite{pizzonia2008}
\nocite{howe1996}
\nocite{rubin1998}
\nocite{cisco2010}
\nocite{cisco2014}
\nocite{juniper2011}
\nocite{checkpoint2008}
%\nocite{guttman1997}
\nocite{bartal1999s}
\nocite{pearce1998}
\nocite{knight2013}
\nocite{libes1995}
\nocite{soule2014}
%\nocite{lazaris2014}
\nocite{jackson2012}
\nocite{bbc2014}
\nocite{levin2013}
\nocite{gutz2012}
\nocite{anderson2014}
\nocite{foster2011}
\nocite{cisco2014b}
\nocite{ranathunga2015T}
\nocite{ranathunga2015V}
\nocite{ranathunga2015P}
\nocite{ranathunga2015W}
\nocite{patrick1995}
\nocite{prakash2015}

%\section*{List of Acronyms}
\addcontentsline{toc}{section}{List of Acronyms}
\begin{acronym}[TDMA]
        \setlength{\itemsep}{-\parsep}
        \acro{IANA}{Internet Assigned Numbers Authority}
        \acro{OOP}{Object Oriented Programming}
        \acro{BNF}{Backus-Naur Form} 
        \acro{MS-SQL}{Microsoft SQL Server}
        \acro{IL}{Intermediate Level}
        \acro{SCADA}{Supervisory Control and Data Acquisition}
        \acro{COTS}{Commodity Off-The-Shelf}
        \acro{TCP}{Transmission Control Protocol}
        \acro{IP}{Internet Protocol}
        \acro{ISA}{International Society for Automation}
        \acro{ASA}{Adaptive Security Appliance}
        \acro{PIX}{Private Internet eXchange}
        \acro{IOS}{IOS}
        \acro{WAN}{Wide Area Network}
        \acro{CLI}{Command Line Interface}
        \acro{DoS}{Denial of Service}         
        \acro{ACLs}{Access Control Lists}
        \acro{ACE}{Access Control Entry}
        \acro{ACEs}{Access Control Entries}
        \acro{CN}{Corporate Network}
        \acro{CZ}{Corporate-Zone}
        \acro{SZ}{SCADA-Zone}
        \acro{MZ}{Management-Zone}
        \acro{FWZ}{Firewall-Zone}
        \acro{IZ}{Internet-Zone}  
        \acro{IDS}{Intrusion Detection Systems} 
        \acro{DMZ}{Demilitarised-Zone}
        \acro{UDP}{User Datagram Protocol}
        \acro{DNS}{Domain Name System}
        \acro{HTTP}{Hypertext Transfer Protocol}
        \acro{HTTPS}{Hypertext Transfer Protocol Secure}
        \acro{FTP}{File Transfer Protocol}
        \acro{NTP}{Network Time Protocol}
        \acro{SSH}{Secure Shell Protocol}
        \acro{SMTP}{Simple Mail Transfer Protocol}
        \acro{SNMP}{Simple Network Management Protocol}
        \acro{ICMP}{Internet Control Message Protocol}
        \acro{DCOM}{Distributed Component Object Model}        
        \acro{NAT}{Network Address Translation}
        \acro{VPN}{Virtual Private Network}
        \acro{VPNs}{Virtual Private Networks}
        \acro{ANSI}{American National Standards Institute}
        \acro{OSI}{Open System Interconnection}
        \acro{SUC}{System Under Consideration}
        \acro{SUCs}{Systems Under Consideration}
        \acro{VM}{Virtual Machine}
        \acro{ACL}{Access Control List}
        \acro{UML}{User Mode Linux}
        \acro{LoC}{Lines of Code}
        \acro{PLCs}{Programmable Logic Controllers}
        \acro{IL}{Intermediate Language }
        
\end{acronym}
\bibliographystyle{abbrv}
\bibliography{myBibliography} 

%%%%%%%%%%%%%%%%%%%%%%% Section 9 %%%%%%%%%%%%%%%%%%%%%%%%%%%%%%%%%%
\appendix
%\section{Appendix}

\begin{figure}[h!]
\caption{High-level policy verification framework using Alloy (partially shown)}
\begin{lstlisting}[caption={}, label=fig:alloy_export2, columns=fullflexible,breaklines=true,basicstyle=\small\fontfamily{cmtt}\selectfont]
abstract sig Service  {
   ip_protocol: some Int,
   source_port: set String,
   dest_port: set String,
   icmp_type: set Int  }
   
abstract sig PolicyRule {
   zone1: one String,
   zone2: one String,
   operator: some Int,
   service: one Service  }
   
// Policy definition
one sig SecurityPolicy { rules: some PolicyRule }

// List of global constraints
fact {
 
 // All defined rules are in the policy to check
 all r: PolicyRule | r in SecurityPolicy.rules

 // Policy rules make up universe of PolicyRule
 SecurityPolicy.rules = PolicyRule

 // A service belongs to at least one PolicyRule
 all s: Service | some r: PolicyRule | s in r.service}
\end{lstlisting}
\end{figure}

\begin{figure*}[t!]
\caption{ForestFirewalls policy description (partially shown).}
\begin{lstlisting}[caption={}, label=fig:suc_policy1,columns=fullflexible,breaklines=
true,basicstyle=\small\fontfamily{cmtt}\selectfont]
// library files
import system.services.iana_services;
import system.services.iana_icmp;

// zone-conduit security topology
load_zone_conduit_model ``zone_conduit.graphml''

// define zone groups
zone_group all_zones {z1,z2,z3,az1,fwz1,fwz2,fwz3}
zone_group scada_zone { z3 }
zone_group corp_zone { z1 }
zone_group internet_zone { z2 }
zone_group all_firewall_zones { fwz1, fwz2, fwz3 }
zone_group all_internal_zones { all_zones \  internet_zone }

// passive mode FTP using custom port numbers
port_group ftp_data_ports { 24500-24600 }
service ftp_data {  protocol=tcp;  tcp.dest_port=ftp_data_ports; }

// service groups using standard port numbers
service_group ftp { iana_services.ftp_control, ftp_data }
service_group web { iana_services.http, iana_services.https }
service_group ping { iana_icmp.icmp_echo, iana_icmp.icmp_echo_reply }
service_group dns { iana_services.dns_tcp, iana_services.dns_udp }
service_group file_transfer { iana_services.http, ftp }

// define security policy
policy_rule file_transfer_rule  { scada_zone -> corp_zone : file_transfer  }

policy_rule ping_rule { corp_zone  <-> scada_zone : ping }

policy_rule dns_rule { scada_zone -> corp_zone : dns  }

policy_rule web_rule  { scada_zone -> corp_zone : web  }
      
rule_group security_policy { file_transfer_rule, ping_rule, dns_rule, web_rule }
                  
// enable policy verification reporting in firewalls
reporting_rule verify_rules{   use_case=verification; 
  granularity.network={zone_or_group={all_zones}};			  
  granularity.policy={rule_or_group={security_policy}};
  granularity.traffic={measurement={counter}; 
  counter_type={connection};};
  granularity.temporal={per_hour};
  granularity.performance={process};}

// define global policy
policy company_policy { security_policy;  verify_rules}
\end{lstlisting}
\end{figure*}

%\subsection{System generated network-level policy (partially shown, comments are denoted by remark).}

%\small
\begin{figure*}[t!]
\caption{System generated network-level policy (partially shown, comments are denoted by remark).}
{\footnotesize
\begin{verbatim}
INFO Vendor neutral network-level ruleset for ACL: acl_2
  remark~enable corp_zone to scada_zone HTTPS traffic (return path)
  permit~tcp~from~10.0.0.16/29~to~10.0.0.0/29~sport~[443]~dport~[`0-65535']~state~ESTABLISHED~log
  permit~tcp~from~10.0.0.16/29~to~10.0.128.4/30~sport~[443]~dport~[`0-65535']~state~ESTABLISHED~log
  remark~enable scada_zone to corp_zone WEB traffic (forward path)
  permit~tcp~from~10.0.0.16/29~to~10.0.0.0/29~sport~[`0-65535']~dport~[443]~state~NEW,ESTABLISHED~log
  permit~tcp~from~10.0.0.16/29~to~10.0.128.4/30~sport~[`0-65535']~dport~[80]~state~NEW,ESTABLISHED~log
  deny~ip~from~any~to~any~sport~~dport~~state~
\end{verbatim}}
\end{figure*}

\end{document}